\documentclass[lettersize,journal]{IEEEtran}
\usepackage{cite}
\usepackage[utf8]{inputenc} 
\usepackage[T1]{fontenc}    
\usepackage[colorlinks]{hyperref}       
\usepackage{url}            
\usepackage{booktabs}       
\usepackage{amsfonts}       
\usepackage{nicefrac}       
\usepackage{microtype}      
\usepackage{xcolor}         

\usepackage{graphicx}
\usepackage{multirow}
\usepackage{makecell}
\usepackage{enumitem}
\usepackage{subcaption}
\usepackage{diagbox}
\usepackage{pifont}
\usepackage{colortbl}
\usepackage{threeparttable}
\usepackage{bm}             
\usepackage{amsmath}
\usepackage{algorithm}
\usepackage{algorithmic}
\usepackage{wrapfig}

\newcommand{\vevo}{Vevo2}

\definecolor{lightgreen}{rgb}{0.88, 1, 0.88}
\definecolor{lightred}{rgb}{1, 0.88, 0.88}
\definecolor{lightyellow}{rgb}{1, 1, 0.88}
\definecolor{lightblue}{rgb}{0.88, 0.95, 1}
\definecolor{lightorange}{rgb}{1, 0.93, 0.88}

\definecolor{yushunred}{rgb}{0.82, 0.18, 0.14}
\definecolor{yushunblue}{rgb}{0.16, 0.33, 0.68}
\definecolor{yushungreen}{rgb}{0.31, 0.65, 0.31}

\newcommand{\colorone}[1]{\textcolor{yushunblue}{{#1}}}
\newcommand{\colortwo}[1]{\textcolor{yushunred}{{#1}}}
\newcommand{\colorthree}[1]{\textcolor{yushungreen}{{#1}}}
\newcommand{\colorpredict}[1]{\textcolor{gray}{{#1}}}

\newcommand{\revisioncolor}{black}
\newenvironment{revision}{%
  \color{\revisioncolor}%
  \arrayrulecolor{\revisioncolor}%
  \captionsetup{font={color=\revisioncolor}}%
  \AtBeginEnvironment{tabular}{\color{\revisioncolor}}%
}{%
  \AtBeginEnvironment{tabular}{\color{black}}%
  \captionsetup{font={color=black}}%
  \arrayrulecolor{black}%
}

\usepackage[most]{tcolorbox}
\newtcolorbox{reviewercomment}{
  colback=gray!5,
  colframe=gray!60,
  boxrule=0.8pt,
  arc=2pt,
  left=6pt,right=6pt,top=6pt,bottom=6pt
}

\begin{document}

\bstctlcite{IEEEexample:BSTcontrol}

\title{\textbf{\textit{Vevo2}}: A Unified and Controllable Framework for \\Speech and Singing Voice Generation}


\author{
Xueyao Zhang,~\IEEEmembership{}
Junan Zhang,~\IEEEmembership{}
Yuancheng Wang,~\IEEEmembership{}
Chaoren Wang,~\IEEEmembership{}\\
Yuanzhe Chen,~\IEEEmembership{}
Dongya Jia,~\IEEEmembership{}
Zhuo Chen,~\IEEEmembership{}
Zhizheng Wu~\IEEEmembership{}


\thanks{Xueyao Zhang, Junan Zhang, Yuancheng Wang, and Chaoren Wang are affiliated with The Chinese University of Hong Kong, Shenzhen. Yuanzhe Chen, Dongya Jia, and Zhuo Chen are affiliated with ByteDance Seed. Zhizheng Wu is affiliated with The Chinese University of Hong Kong, Shenzhen, Shenzhen Loop Area Institute, City University of Macau, and Amphion Technology Co., Ltd. Zhizheng Wu is the corresponding author. e-mail: \href{mailto:wuzhizheng@cuhk.edu.cn}{wuzhizheng@cuhk.edu.cn}.}
}

\markboth{Journal of \LaTeX\ Class Files,~Vol.~14, No.~8, August~2021}%
{Shell \MakeLowercase{\textit{et al.}}: A Sample Article Using IEEEtran.cls for IEEE Journals}


\maketitle

\begin{abstract}
Controllable human voice generation, particularly for expressive domains like singing, remains a significant challenge. This paper introduces \textbf{\textit{\vevo{}}}, a unified framework for controllable speech and singing voice generation. To tackle issues like the scarcity of annotated singing data and to enable flexible controllability, \vevo{} introduces two audio tokenizers: (1) a unified music-notation-free \textit{prosody tokenizer} that captures prosody and melody from speech, singing, and even instrumental sounds, and (2) a unified \textit{content-style tokenizer} that encodes linguistic content, prosody, and style for both speech and singing, while enabling timbre disentanglement. \vevo{} consists of an auto-regressive content-style modeling stage, which aims to enable controllability over text, prosody, and style, as well as a flow-matching acoustic modeling stage that allows for timbre control. Particularly, during the speech-singing joint training of the AR model, we propose both \textit{explicit} and \textit{implicit} prosody learning strategies to bridge speech and singing voice. Moreover, to further enhance the \vevo{}’s ability to follow text and prosody, we design a multi-objective post-training task that integrates both intelligibility and prosody similarity alignment. Experimental results show that the unified modeling in \vevo{} brings mutual benefits to both speech and singing voice generation. Additionally, \vevo{}’s effectiveness across a wide range of synthesis, conversion, and editing tasks for both speech and singing further demonstrates its strong generalization ability and versatility. Audio samples are are available at \url{https://versasinger.github.io/}.
\end{abstract}


\section{Introduction}
Controllable generation of human voice is an important research topic in the field of audio generation~\cite{tts-book-tanxu,xietianxin-survey}. In recent years, remarkable progress has been achieved in speech generation, particularly in zero-shot text-to-speech (TTS)~\cite{naturalspeech3,seedtts,maskgct,cosyvoice}, largely propelled by the availability of large-scale training corpora. However, for more challenging domains such as singing voice—which can be regarded as a highly expressive form of speech with strictly constrained prosody (e.g., \textit{melody})\footnote{This paper adopts a definition in which \textit{prosody} refers to the suprasegmental attributes of human voice, including pitch, duration, and loudness. We regard \textit{melody}—which primarily involves pitch and duration (with rhythm viewed as a temporal organization of duration)—as a specific and more constrained form of prosody.}~\cite{singingvoice-science}—achieving both high generation quality and controllability remains challenging~\cite{seedmusic,yue}.

This paper focuses on unifying controllable speech and singing voice generation. Our motivation is based on the hypothesis that learning for speech and singing voice can be mutually beneficial through unified modeling. Specifically, (1) \textbf{\textit{Speech $\rightarrow$ Singing Voice}}: the abundance of speech data can supplement the typically smaller singing voice datasets~\cite{emilia,gtsinger}, potentially improving the overall quality of generated singing voice~\cite{yue}. (2) \textbf{\textit{Singing Voice $\rightarrow$ Speech}}: the inherent expressiveness of singing voice, when incorporated into a unified learning framework, can enhance expressive speech generation~\cite{yicheng-trans}. Besides, since singing voices inherently contain structured melodic patterns, learning to model them helps the system enhance its ability to follow melody, thereby improving overall prosody-following capability. This capability is particularly beneficial for tasks such as prosody-preserved speech editing (Section~\ref{sec:expt-controllability-text-prosody}).

Several recent studies have attempted to unify speech and singing voice generation. For instance, UniSyn~\cite{unisyn} integrates TTS and Singing Voice Synthesis (SVS) through a task identifier that switches between generation modes, while UniAudio~\cite{uniaudio} extends this concept to a broader range of audio generation tasks via task-conditioned sequence modeling. Although these approaches represent valuable progress, they still face two fundamental challenges. \textbf{First}, the existing corpora for singing voice generation rely heavily on expert annotations~\cite{singing-expression,diffsinger,tcsinger,spsinger}, such as detailed \textit{music notation} with precise alignment between lyrics and musical notes~\cite{m4singer,singstyle111,gtsinger}. These datasets are scarce, and their unique data organization is inherently inconsistent with large-scale speech corpora, thereby limiting scalability and unified training. \textbf{Second}, achieving versatile controllability within such a unified system remains nontrivial—such as the independent manipulation of text (or lyrics), prosody (or melody), style (e.g., accent, emotion, or \text{singing techniques}~\cite{vocalset,gtsinger}), and timbre (i.e., speaker identity) within a single framework.

\begin{figure*}[t]
    \centering
    \begin{subfigure}[b]{0.9\textwidth}
        \centering
        \includegraphics[width=\textwidth]{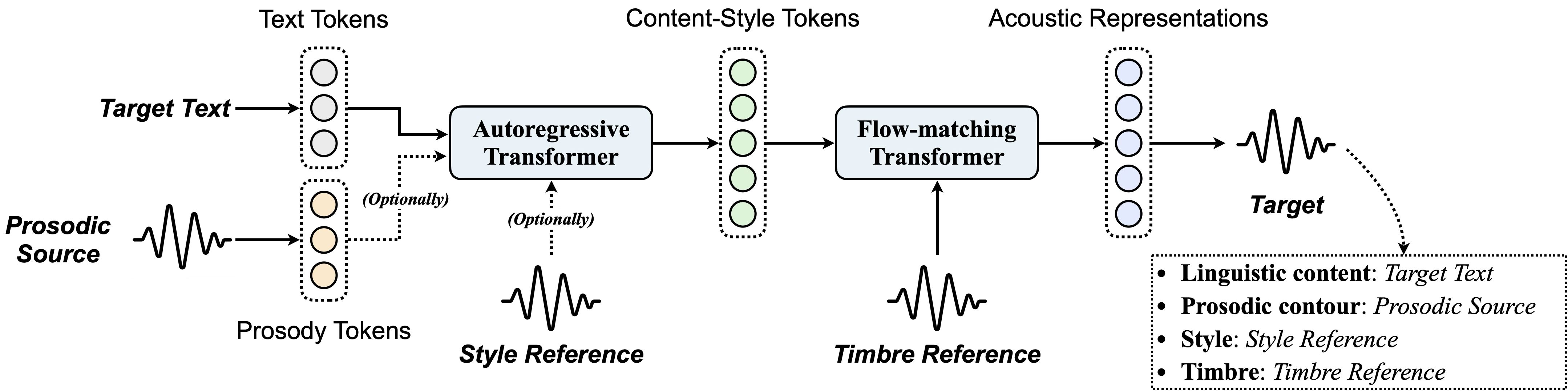}
    \end{subfigure}

    \vspace{1mm}
    
    \begin{subfigure}[b]{0.9\textwidth}
        \centering
        \resizebox{\textwidth}{!}{%
\begin{threeparttable}
    \begin{tabular}{c|c|c|c}
    \toprule
    \textbf{Task} & \textbf{Input} & \textbf{Auto-Regressive Stage} & \textbf{Flow-Matching Stage} \\
    \midrule
    \textbf{Text to Speech} & \colorone{\textbf{text}}, \colortwo{\textbf{reference wav}} &[\colortwo{$\pmb{\bm{t}}$}, \colorone{$\pmb{\bm{t}}$}, \colortwo{$\pmb{\bm{cs}}$}] $\rightarrow$ [\colorpredict{$\pmb{\bm{\hat{cs}}}$}] & \multicolumn{1}{c}{\multirow{8}{*}{[\colortwo{$\pmb{\bm{cs}}$}, \colortwo{$\pmb{\bm{mel}}$}, \colorpredict{$\pmb{\bm{\hat{cs}}}$}] $\rightarrow$ [\colorpredict{$\pmb{\bm{\hat{mel}}}$}]}} \\
    \cmidrule(lr){1-1} \cmidrule(lr){2-2} \cmidrule(lr){3-3}
    \textbf{Singing Voice Synthesis} & \colorone{\textbf{text}}, \colorthree{\textbf{midi}}, \colortwo{\textbf{reference wav}} & \multirow{1}{*}{[\colortwo{$\pmb{\bm{t}}$}, \colorone{$\pmb{\bm{t}}$}, \colortwo{$\pmb{\bm{p}}$}, \colorthree{$\pmb{\bm{p}}$}, \colortwo{$\pmb{\bm{cs}}$}] $\rightarrow$ [\colorpredict{$\pmb{\bm{\hat{cs}}}$}]} & \\
    \cmidrule(lr){1-1} \cmidrule(lr){2-2} \cmidrule(lr){3-3}
    \makecell[c]{\textbf{Voice Conversion} \textit{(style-converted)}} & \multirow{2}{*}{\colorone{\textbf{source wav}}, \colortwo{\textbf{reference wav}}} & [\colortwo{$\pmb{\bm{t}}$}, \colorone{$\pmb{\bm{t}}$}, \colortwo{$\pmb{\bm{cs}}$}] $\rightarrow$ [\colorpredict{$\pmb{\bm{\hat{cs}}}$}] &  \\
    \cmidrule(lr){1-1} \cmidrule(lr){3-3}
    \makecell[c]{\textbf{Singing Voice Conversion} \textit{(style-converted)}} &  & [\colortwo{$\pmb{\bm{t}}$}, \colorone{$\pmb{\bm{t}}$}, \colortwo{$\pmb{\bm{p}}$}, \colorone{$\pmb{\bm{p}}$}, \colortwo{$\pmb{\bm{cs}}$}] $\rightarrow$ [\colorpredict{$\pmb{\bm{\hat{cs}}}$}] & \\
    \cmidrule(lr){1-1} \cmidrule(lr){2-2} \cmidrule(lr){3-3}
    \textbf{Speech Editing} & \multirow{2}{*}{\colorone{\textbf{edited text}}, \colortwo{\textbf{raw wav}}} & \multirow{2}{*}{[\colortwo{$\pmb{\bm{t}}$}, \colorone{$\pmb{\bm{t}}$}, \colortwo{$\pmb{\bm{p}}$}, \colorone{$\pmb{\bm{p}}$}, \colortwo{$\pmb{\bm{cs}}$}] $\rightarrow$ [\colorpredict{$\pmb{\bm{\hat{cs}}}$}]} & \\
    \cmidrule(lr){1-1}
    \textbf{Singing Lyric Editing} & & & \\
    \cmidrule(lr){1-1} \cmidrule(lr){2-2} \cmidrule(lr){3-3}
    \textbf{Humming to Singing} & \colorone{\textbf{text}}, \colorthree{\textbf{humming wav}}, \colortwo{\textbf{singer wav}} & \multirow{2}{*}{[\colorone{$\pmb{\bm{t}}$}, \colorthree{$\pmb{\bm{p}}$}] $\rightarrow$ [\colorpredict{$\pmb{\bm{\hat{cs}}}$}]} & \\
    \cmidrule(lr){1-1} \cmidrule(lr){2-2}
    \textbf{Instrument to Singing} & \colorone{\textbf{text}}, \colorthree{\textbf{instrument wav}}, \colortwo{\textbf{singer wav}} & & \\
    \midrule
    \textbf{Emotion Conversion} & \multirow{4}{*}{\colorone{\textbf{source wav}}, \colortwo{\textbf{style wav}}} & \multirow{3}{*}{[\colortwo{$\pmb{\bm{t}}$}, \colorone{$\pmb{\bm{t}}$}, \colortwo{$\pmb{\bm{cs}}$}] $\rightarrow$ [\colorpredict{$\pmb{\bm{\hat{cs}}}$}]} & \multirow{4}{*}{[\colorone{$\pmb{\bm{cs}}$}, \colorone{$\pmb{\bm{mel}}$}, \colorpredict{$\pmb{\bm{\hat{cs}}}$}] $\rightarrow$ [\colorpredict{$\pmb{\bm{\hat{mel}}}$}]} \\
    \cmidrule(lr){1-1}
    \textbf{Accent Conversion} & & & \\
    \cmidrule(lr){1-1}
    \textbf{Whisper-to-Normal Conversion} & & & \\
    \cmidrule(lr){1-1} \cmidrule(lr){3-3}
    \textbf{Singing Style Conversion} & & [\colortwo{$\pmb{\bm{t}}$}, \colorone{$\pmb{\bm{t}}$}, \colortwo{$\pmb{\bm{p}}$}, \colorone{$\pmb{\bm{p}}$}, \colortwo{$\pmb{\bm{cs}}$}] $\rightarrow$ [\colorpredict{$\pmb{\bm{\hat{cs}}}$}] & \\
    \midrule
    \makecell[c]{\textbf{Voice Conversion} \textit{(style-preserved)}} & \multirow{2}{*}{\colorone{\textbf{source wav}}, \colortwo{\textbf{reference wav}}} & \multirow{2}{*}{-} &  \multirow{2}{*}{[\colortwo{$\pmb{\bm{cs}}$}, \colortwo{$\pmb{\bm{mel}}$}, \colorone{$\pmb{\bm{cs}}$}] $\rightarrow$ [\colorpredict{$\pmb{\bm{\hat{mel}}}$}]} \\
    \cmidrule(lr){1-1}
    \makecell[c]{\textbf{Singing Voice Conversion} \textit{(style-preserved)}} & & & \\
    \bottomrule
    \end{tabular}
    \begin{tablenotes}
        \footnotesize{\item[*] We denote the text tokens, prosody tokens, content-style tokens, and acoustic representations (specifically, Mel spectrogram) as $\pmb{\bm{t}}$, $\pmb{\bm{p}}$, $\pmb{\bm{cs}}$, and $\pmb{\bm{mel}}$, respectively. We use \colorone{\textbf{blue}}, \colortwo{\textbf{red}}, and \colorthree{\textbf{green}} to indicate the source inputs of these representations. \colorpredict{$\pmb{\bm{\hat{cs}}}$} and \colorpredict{$\pmb{\bm{\hat{mel}}}$} represent the model-predicted outputs. For the singing voice synthesis (SVS) task, we render MIDI notations into instrumental sounds and then extract their prosody tokens (Section~\ref{sec:expt-controllability-text-prosody}).} 
    \end{tablenotes}
\end{threeparttable}
}
    \end{subfigure}
    \caption{\vevo{} inference pipeline for versatile synthesis, conversion, and editing tasks.}
    \label{fig:vevo2-pipeline}
\end{figure*}

\begin{figure}[t]
    \centering
    \begin{subfigure}[b]{0.4\columnwidth}
        \centering
        \includegraphics[width=\textwidth]{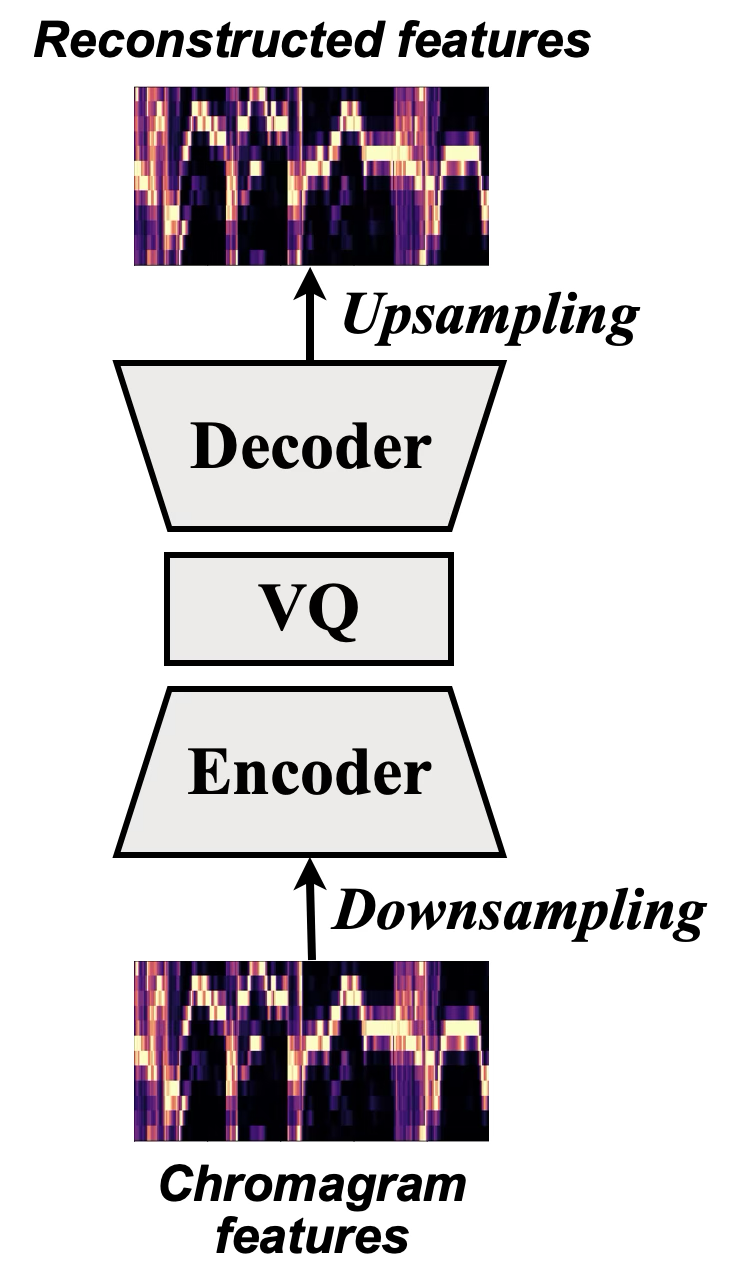}
        \caption{Prosody Tokenizer}
        \label{fig:vevo2-tokenizer-prosody}
    \end{subfigure}
    \hspace{4mm}
    \begin{subfigure}[b]{0.41\columnwidth}
        \centering
        \includegraphics[width=\textwidth]{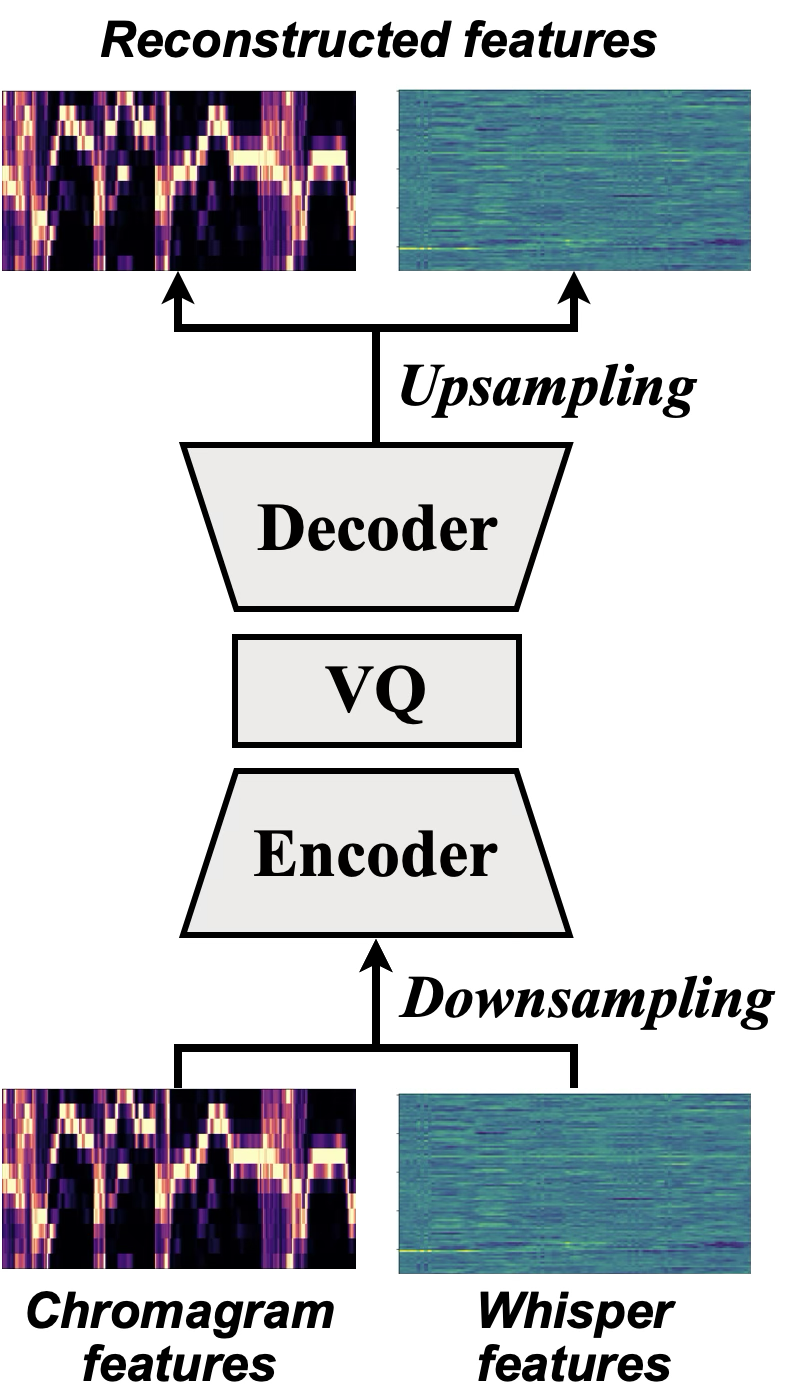}
        \caption{Content-Style Tokenizer}
        \label{fig:vevo2-tokenizer-contentstyle}
    \end{subfigure}
\caption{Unified speech and singing tokenizers of \vevo{}.}
\label{fig:vevo2-tokenizer}
\end{figure}

To address these challenges, this paper introduces \textit{\textbf{\vevo{}}}, a \underline{ve}rsatile and controllable \underline{vo}ice generation framework designed to unify speech and singing voice. \vevo{} adopts a two-stage architecture, similar to classic zero-shot speech generation systems~\cite{seedtts,cosyvoice,vevo} (Figure~\ref{fig:vevo2-pipeline}): \textit{\textbf{(1) Auto-Regressive (AR) Content-Style Modeling Stage}}: given a target text and a prosodic source as input, we utilize an AR transformer to generate the content-style tokens, which is prompted by a \textit{style reference}. \textit{\textbf{(2) Flow-Matching (FM) Acoustic Modeling Stage}}: given content-style tokens as input, we employ an FM transformer to produce acoustic representations (specifically, Mel spectrograms in our study), which is prompted by a \textit{timbre reference}. To achieve the desired unification and controllability, \vevo{} introduces the following three key method designs:

\textbf{1. Unified Speech and Singing Voice Tokenizers.} 
We propose two distinct audio tokenizers: a unified \text{prosody tokenizer} and a unified \text{content-style tokenizer}. \textit{\textbf{(1) Prosody Tokenizer}}: A single-codebook VQ-VAE tokenizer~\cite{vq-vae,rvqgan} operating at 6.25~Hz (56.25~bps), trained to reconstruct the \textit{chromagram}~\cite{meinard-book} of raw audio (Figure~\ref{fig:vevo2-tokenizer-prosody}). 
This tokenizer unifies the modeling of both melody in singing voices and prosody in speech, and can directly extract tokens from the chromagram of raw audio, eliminating the need for expert-annotated music notation and thus greatly improving scalability. 
Moreover, due to the chromagram's strong generalization ability across various types of audio data~\cite{meinard-book,musicgen}, this tokenizer can even capture the melody of non-human sounds such as instrumental music (Section~\ref{sec:ablation-grpo}), thereby enabling unique music generation applications, such as humming-to-singing and instrument-to-singing. 
\textit{\textbf{(2) Content-Style Tokenizer}}: It employs a similar architecture to the prosody tokenizer, but operates at 12.5~Hz (175~bps) and is trained to reconstruct both chromagram features and Whisper hidden features~\cite{whisper} (Figure~\ref{fig:vevo2-tokenizer-contentstyle}). 
It effectively captures linguistic content, melody, and style in both speech and singing voice, while also achieving robust timbre disentanglement (Table~\ref{tab:expt-svc}). 

\textbf{2. Speech-Singing Joint Training.} 
To address the joint training of speech and singing voice, we design two different training strategies during the content-style modeling stage:
\textit{\textbf{(1) Explicit Prosody Learning (EPL)}}, where prosody tokens are provided as explicit input, allowing the model to predict content-style tokens from both text and prosody tokens; and \textit{\textbf{(2) Implicit Prosody Learning (IPL)}}, where prosody tokens are not used as input, and content-style tokens are predicted solely from text tokens. 
By randomly applying both EPL and IPL to all voice data, the model learns prosody in a more unified manner, effectively bridging the gap between speech and singing characteristics. 

\textbf{3. Multi-objective Alignment.} 
To further improve the text- and prosody-following abilities of the pre-trained model, we introduce a multi-objective post-training task.
We propose a composite reward that incorporates both intelligibility and prosody similarity, and adopt the Group Relative Policy Optimization (GRPO) algorithm~\cite{grpo} for training. 
Experimental results demonstrate that this post-training significantly enhances the model’s controllability with respect to text and melody, and also improves its generalization to out-of-pretraining-distribution data, such as instrumental sounds.

The contributions of this paper are summarized as follows:
\begin{itemize}[itemsep=0ex,leftmargin=3ex]
    \item We propose two audio tokenizers: (1) a unified music-notation-free prosody tokenizer capable of encoding prosody and melody from speech, singing voice, and even instrumental sounds; and (2) a unified timbre-disentangled content-style tokenizer that models linguistic content, prosody, and style for both speech and singing voice.
    \item We introduce \vevo{}, a unified and controllable framework for both speech and singing voice generation. Built upon the proposed speech-singing joint training, \vevo{} can be applied to versatile voice synthesis, conversion, and editing tasks by flexibly combining text, prosodic source, and style and timbre references during inference.
    \item We propose a multi-objective post-training task for controllable voice generation. By optimizing both intelligibility and prosody similarity rewards with a GRPO-based algorithm, \vevo{} achieves greater stability and improved overall performance across a variety of tasks.
\end{itemize}

We will release code and model checkpoints of this study at Amphion\footnote{\url{https://github.com/open-mmlab/Amphion}}~\cite{amphion,amphion_v0.2}.

\section{Related Work}

\textbf{Controllable Voice Generation}\quad
Controllable speech and singing voice generation has consistently remained a significant research focus. (1) In the speech domain, existing works explore control over multiple attributes: \textit{linguistic content}, which is fundamental to TTS~\cite{tts-book-tanxu,xietianxin-survey} and speech editing tasks~\cite{voicebox,voicecraft,ssrspeech,f5tts}; \textit{timbre}, which is central to VC tasks~\cite{ppg-vc,autovc,naturalspeech3,voiceshop,vevo,seedvc}; and \textit{style}, which encompasses accent~\cite{parallel-ac-zhaoguanlong21,convertandspeak,vevo}, emotion~\cite{esd,pavits,vevo}, whisper~\cite{whisper-to-normal-lstm}, and a broader natural language description~\cite{parler-tts,controlspeech,paraspeechcaps}.
(2) For singing voice, beyond controlling \textit{lyrics}, \textit{timbre}, and \textit{style}—as explored in SVS~\cite{diffsinger,visinger,tcsinger,stylesinger,techsinger}, SVC~\cite{svcc-2023,svcc-vits-ziqian,neucosvc,amphion-svc}, and singing lyric editing tasks~\cite{editsinger,unisinger,songeditor}—\textit{melody} control holds particular significance due to the inherent musicality of singing voices~\cite{musicgen,seedmusic,yue}.
(3) From a unified modeling perspective, previous research has attempted to integrate various speech generation tasks such as TTS and VC~\cite{vevo,cosyvoice,metis}, or different singing voice generation tasks like SVS and SVC~\cite{unisinger,everyone-can-sing}. However, models capable of unifying both speech and singing voice domains remain scarce~\cite{uniaudio,unisyn}. Moreover, the potential benefits of joint speech and singing pre-training, along with the versatile control mechanisms for multiple voice attributes within a unified system, remain largely unexplored.

\textbf{Speech Tokenizer}\quad
Recent research has emphasized the disentangled speech representation learning, aiming to separately model the factorized speech attributes to reduce learning complexity~\cite{megatts,naturalspeech3} and achieve more precise controllability~\cite{speech-resynthesis-interspeech21,naturalspeech3,vevo}. Common speech tokenizers include linguistic content tokenizers~\cite{hubert,vevo,syllablelm}, content-style tokenizers~\cite{audiolm,seedtts,maskgct,cosyvoice,vevo}, and prosody tokenizers~\cite{megatts,naturalspeech3}. However, in pursuing a unified controllable model for both speech and singing generation, existing speech tokenizers exhibit several limitations: As these tokenizers predominantly focus on the speech domain, their effectiveness in handling singing voice data remains unexplored. The design of a unified prosody tokenizer capable of handling both modalities, along with enhancing the melody encoding capabilities of content-style tokenizers~\cite{amphion-svc}, remains an important issue for future exploration.

\textbf{Post-training for Audio Generation}\quad
Alignment via post-training has demonstrated its effectiveness in the generation of text~\cite{instructgpt,RLHF-anthoropic}, vision~\cite{imagereward,tldr-perturb-vision}, speech~\cite{speechalign,seedtts,cosyvoice2}, music~\cite{musicrl}, and sound effects~\cite{tango2,baton}. In speech generation, existing works have employed preference alignment to enhance multiple aspects of speech, including intelligibility~\cite{seedtts,cosyvoice2,intp}, speaker similarity~\cite{seedtts,cosyvoice2}, emotion controllability~\cite{seedtts,emo-dpo}, and overall quality~\cite{speechalign,uno,fpo}. For developing a unified speech and singing voice generation model, the importance of specific capability elicitation through post-training becomes particularly crucial, given the expanded range of capabilities injected into the pre-trained model. However, the design of appropriate alignment objectives for unified speech and singing voice generation remains an open research problem.

\section{Methodology}\label{sec:method}


The overall architecture of \vevo{} is a two-stage, versatile voice generation framework that unifies both speech and singing voice (Figure~\ref{fig:vevo2-pipeline}). 
During inference, by flexibly combining different input conditions—such as text, prosodic source, style reference, and timbre reference—\vevo{} supports a wide range of voice-related synthesis, conversion, and editing tasks (Section~\ref{sec:inference-time-control}). Specifically, \vevo{} first encodes linguistic, stylistic, and prosodic cues into discrete content-style tokens through an AR transformer, and then reconstructs Mel spectrograms via an FM transformer. 
These spectrograms are subsequently converted into waveforms using a Vocos-based vocoder~\cite{vocos}. To achieve the unified speech-singing voice generation and versatile controllability, \vevo{} is built upon several key methodological designs, which are detailed in the following sections.


    

\begin{figure*}[t]
    \centering
    \begin{subfigure}[b]{0.9\textwidth}
        \centering
        \includegraphics[width=\textwidth]{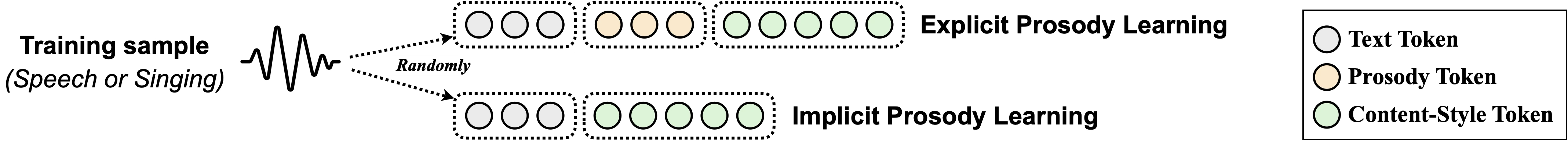}
    \end{subfigure}
\caption{Speech-Singing Joint Training with Explicit Prosody Learning (EPL) and Implicit Prosody Learning (IPL). We perform the next token prediction only on the sequence of content-style tokens (see Section~\ref{sec:pre-training} for more details).}
\label{fig:vevo2-epl-ipl}
\end{figure*}


\subsection{Unified Speech and Singing Voice Tokenizers}\label{sec:tokenizer}

\textbf{Prosody Tokenizer}\quad
In order to model both the prosody of speech and the melody of singing voice using a unified method, we choose \textit{chromagram feature}~\cite{meinard-book} and design a VQ-VAE tokenizer~\cite{vq-vae,rvqgan} based on it. Some previous studies of speech generation tried to use fundamental frequency (F0)~\cite{speech-resynthesis-interspeech21,naturalspeech3} or the low-frequency band of the mel spectrogram~\cite{megatts} to build a prosody tokenizer. The reasons why we adopt the chromagram are as follows:
\begin{enumerate}[itemsep=0ex,leftmargin=3ex]
    \item \textbf{Octave-free, thus friendly for unification}: There is a significant distribution gap of F0 between speech and singing voice, such as the F0 range of singing voice is much wider~\cite{my-svc-paper-workshop}. Chromagram removes octave information between singing voice and speech, making it easier to capture common prosodic patterns across both domains.
    \item \textbf{Notation-free, thus friendly for scaling up}: Chromagram effectively represents high-level musical and singing voice attributes such as melody and harmony~\cite{meinard-book}, and can be automatically extracted from audio without manual transcription or music notation, making it suitable for large-scale data processing.
    \item \textbf{Robust for diverse audio domains}: As a statistical distribution of spectral energy, chromagram is less sensitive to noise and pitch detection errors~\cite{yin}, resulting in more stable performance. In addition, chromagram provides strong representations for both human voice and non-human sounds, such as instrumental music~\cite{meinard-book,musicgen}. These characteristics enhance the flexibility of melody control in downstream music generation tasks, such as enabling instrument-to-sing applications (Section~\ref{sec:ablation-grpo}).
\end{enumerate}



\textcolor{\revisioncolor}{Although a continuous chromagram can be directly used as a conditioning signal, we adopt VQ quantization to obtain compact \emph{discrete} prosody tokens that are better suited for our sequence modeling pipeline and more robust to acoustic variations (see Appendix for an empirical analysis).} The prosody tokenizer consists of five main components (Figure~\ref{fig:vevo2-tokenizer-prosody}): Downsampling Layer (CNN-based), Encoder, Vector Quantization (VQ), Decoder, and Upsampling Layer (CNN-based). Formally, given the codebook $\bm{E} = [\bm{e}_1, \bm{e}_2, \dots, \bm{e}_K]$ with vocabulary size $K$, and taking chromagram features $\bm{x}$ as input, the reconstructed output $\hat{\bm{x}}$ is obtained after passing through these five modules:
\begin{equation}
\begin{split}
    \bm{z}_e(\bm{x}) &= \text{Encoder}(\text{Downsample}(\bm{x})), \\
    \bm{z}_q(\bm{x}) &= \bm{e}_k,~\text{where}~k = \arg\min_j \|\bm{z}_e(\bm{x}) - \bm{e}_j\|_2, \\
    \hat{\bm{x}} &= \text{Upsample}(\text{Decoder}(\bm{z}_q(\bm{x}))),
\end{split}
\end{equation}
where $\bm{z}_q(\bm{x})$ is the quantized representation (i.e., token) of $\bm{z}_e(\bm{x})$ after VQ. The loss function consists of the reconstruction loss (whose weight is $\lambda$) and quantization loss (whose weight is $\beta$):
\begin{equation}
    \mathcal{L} = \lambda \|\bm{x} - \hat{\bm{x}}\|^2_2 + \beta \|\bm{z}_e(\bm{x}) - \bm{z}_q(\bm{x})\|^2_2.
\end{equation}
We follow~\cite{vqgan-cv,rvqgan,maskgct} and adopt factorized (i.e., a low-dimensional code-lookup space) and L2-normalized strategies for VQ learning. We extract 50 Hz chromagram features from raw audio and apply an 8x downsampling, resulting in a prosody tokenizer operating at 6.25 Hz (56.25 bps).


\textbf{Content-Style Tokenizer}\quad
We follow the main ideas from recent works in controllable speech generation~\cite{vevo,cosyvoice}, and aim to develop a content-style tokenizer that can encode linguistic content, prosody, and style, while also achieving timbre disentanglement. Specially, to unify both speech and singing voice, beyond encoding linguistic content and style, as in typical speech generation tasks~\cite{vevo,cosyvoice,seedtts,maskgct,glm4-voice}, we also require the tokenizer to capture melody and fine-grained style information (e.g., F0 and energy details) in singing voice. Motivated by this, when designing our VQ-VAE tokenizer, we not only use hidden features from automatic speech recognition (ASR) models—specifically, from Whisper~\cite{whisper}—as reconstruction objectives, but also incorporate chromagram features as an additional reconstruction objective. This compensates for the limitations of Whisper features in modeling melody in singing voice~\cite{amphion-svc}. Furthermore, to facilitate the downstream content-style modeling stage, we design the content-style tokenizer to operate at a relatively low frame rate (specifically, 12.5 Hz in our study). This reduces the sequence length for the AR model and thus alleviates its learning burden.

The architecture of the content-style tokenizer is similar to the proposed prosody tokenizer, which is displayed in Figure~\ref{fig:vevo2-tokenizer-contentstyle}. The only difference lies in the input and reconstruction objectives incorporate both chromagram features and Whisper features. We extract 50 Hz chromagram features and 50 Hz Whisper features (which is the encoder output of the pre-trained Whisper model~\cite{whisper}) and apply a 4x downsampling, resulting in a content-style tokenizer operating at 12.5 Hz (175 bps).


\subsection{Speech-Singing Joint Training}\label{sec:pre-training}

To conduct the joint training of speech and singing voice, an essential question lies in how to unify the generation process of speech and singing within a single framework. For the acoustic modeling stage, once the proposed unified content-style tokenizer is obtained, this part becomes relatively straightforward, as both speech and singing voice can be formulated into unified content-style tokens and decoded into Mel spectrograms under the guidance of a timbre reference. 
However, we observe that the main challenge of speech-singing joint training arises in the content-style modeling stage, where the generation characteristics of speech and singing differ substantially. For instance, speech generation like TTS usually takes only text (and the reference) as input and does not use an explicit control over prosody~\cite{tts-book-tanxu}, while singing voice generation often demands direct control over melody~\cite{singing-expression,diffsinger,tcsinger,spsinger}. 

\begin{figure*}[t]
    \centering
    \includegraphics[width=\textwidth]{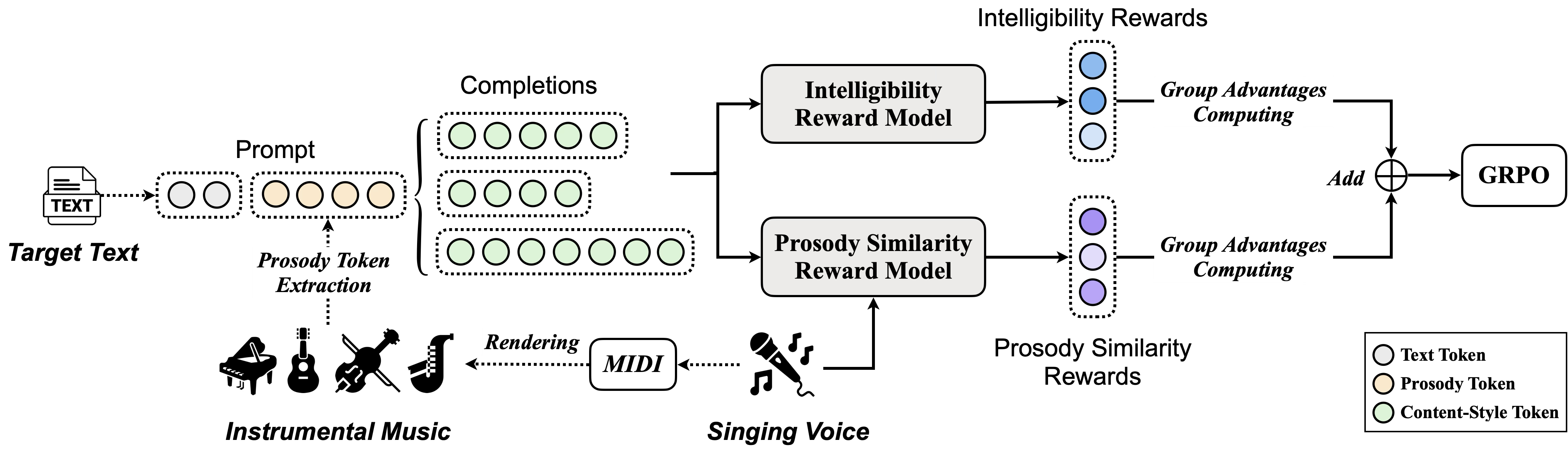}
    \caption{Multi-objective alignment for both intelligibility and prosody similarity. This figure demonstrates how we utilize instrumental music as prosody prompts during post-training of \vevo{}.}
    \vspace{-3mm}
    \label{fig:vevo2-grpo}
\end{figure*}

Motivated by this observation, we focus on the unified prosody learning and propose two different training strategies to bridge speech and singing voice generation during the content-style modeling stage (Figure~\ref{fig:vevo2-epl-ipl}): 
\textbf{(1) Implicit Prosody Learning (IPL)}: the AR model generates content-style tokens solely from text input, implicitly inferring prosodic information through in-context learning. This strategy is commonly adopted in zero-shot TTS systems~\cite{audiolm,seedtts,cosyvoice,vevo,maskgct}. 
\textbf{(2) Explicit Prosody Learning (EPL)}: in addition to the text input, prosody tokens are explicitly provided, allowing the AR model to predict content-style tokens conditioned on both text and prosody information.
Notably, during the joint training, we do not restrict IPL to speech data and EPL to singing voice data. Instead, we treat all samples equally: for each training sample, we randomly choose between IPL and EPL with equal probability. We assume that this design better unifies the training patterns for speech and singing voice. 

Given a speech or singing voice sample ${u}$, we denote the prosody and content-style tokens as $\bm{Q}_{p} ({u})$ and $\bm{Q}_{cs} ({u})$, respectively. The input sequences to the transformer for IPL and EPL are shown in Equation (\ref{eq:ipl}) and (\ref{eq:epl}), where $\bm{I}_{ipl}$ and $\bm{I}_{epl}$ represent the text instruction prefixes for IPL and EPL, $\bm{T}$ denotes the text of ${u}$, and $\langle |\cdot| \rangle$ indicates special tokens:
\begin{equation}
    \textbf{IPL}: [\bm{I}_{ipl}, \bm{T}, \langle |\text{start-of-cs}| \rangle, \bm{Q}_{cs} ({u}),\langle |\text{end-of-cs}| \rangle],
\label{eq:ipl}
\end{equation}
\begin{equation}
\begin{split}
    \textbf{EPL}: [\bm{I}_{epl}, \bm{T},  \langle |\text{start-of-p}| \rangle, \bm{Q}_{p} ({u}), \langle |\text{end-of-p}| \rangle,\\ \langle |\text{start-of-cs}| \rangle, \bm{Q}_{cs} ({u}),\langle |\text{end-of-cs}| \rangle].
\end{split}
\label{eq:epl}
\end{equation}
Specifically, the text instruction prefixes for IPL ($\bm{I}_{ipl}$) and EPL ($\bm{I}_{epl}$) are defined as: \textit{"User will provide you with a text. Please vocalize it with natural expression."} and \textit{"User will provide you with a text. Please first generate a good prosodic instruction, then vocalize the text based on it."} respectively. We perform next token prediction only on the final segment $[\langle |\text{start-of-cs}| \rangle, \bm{Q}_{cs} ({u}),\langle |\text{end-of-cs}| \rangle]$ for both IPL and EPL.

During the acoustic modeling stage, we employ an FM transformer to convert the content-style tokens into Mel spectrograms, conditioned on a timbre reference. 
This process follows the general paradigm used in speech generation studies~\cite{seedmusic,cosyvoice,vevo,glm4-iclr}, 
with the distinction that our approach leverages the proposed unified content-style tokenizer and is jointly trained on both speech and singing voice data. After obtaining the Mel spectrograms, we use a Vocos-based vocoder~\cite{vocos}, which is also trained on both speech and singing voice, to reconstruct the final waveform.


\subsection{Multi-Objective Alignment}\label{sec:post-training}

After pre-training, we observe that the AR model exhibits good versatility across diverse tasks (Section~\ref{sec:results}). Nevertheless, its stability remains suboptimal, particularly in its fundamental capabilities of \textit{text-} and \textit{prosody-following}, which can be further elicited and enhanced. Furthermore, in downstream applications, the model may encounter out-of-pretraining data such as instrumental sounds. For instance, in the context of SVS, the model needs to accept prosody tokens extracted from MIDI-rendered sounds for melody control (Figure~\ref{fig:vevo2-pipeline}). 

Motivated by the above, this study further employs post-training to enhance the model’s controllability for text and prosody, as well as to improve its robustness to diverse audio data. We propose a multi-objective post-training task that jointly aligns both intelligibility and prosody similarity (Figure~\ref{fig:vevo2-grpo}). We believe that this joint alignment helps prevent the model from over-optimizing a single objective (as evidenced in Section~\ref{sec:ablation-grpo}). 


\textbf{Intelligibility Reward}\quad
To align intelligibility, we adopt a recent zero-shot TTS intelligibility preference corpus, INTP~\cite{intp}, which contains diverse intelligibility preference pairs constructed through comparisons between multiple models as well as manually designed intelligibility perturbations, to train an intelligibility reward model. Specifically, we employ the commonly used Bradley-Terry (BT) approach~\cite{bt-rm,instructgpt,RLHF-anthoropic} for reward model training. Given an intelligibility preference dataset $\mathcal{D} = \{(t, a^w, a^l)\}$, where $a^w$ and $a^l$ represent the positive and negative audio outputs with respect to the target text $t$, we parametrize an intelligibility reward model $r_\phi(t, a)$, treat the problem as binary classification, and optimize the negative log-likelihood loss:
\begin{revision}
\begin{equation}
    \mathcal{L}_R(r_\phi, \mathcal{D}) = -\mathbb{E}_{(t, a^w, a^l) \sim \mathcal{D}} \left[ \log \sigma \left( r_\phi(t, a^w) - r_\phi(t, a^l) \right) \right],
\label{eq:btrm}
\end{equation}
\end{revision}
where $r_\phi(t, a)$ is initialized from the pre-trained model. We follow the pre-training and use both EPL and IPL for training this reward model. During inference, we can input it a $[\bm{T}, \bm{Q}_{cs}]$ or $[\bm{T}, \bm{Q}_{p}, \bm{Q}_{cs}]$ sequence to obtain an intelligibility reward. 


\textbf{Prosody Similarity Reward}\quad
To achieve prosodic alignment, we establish the following protocol to evaluate the policy model's prosody-following capability: (1) Suppose we have a singing voice dataset with MIDI annotations. Given a singing voice sample ${u}$, we use its MIDI-rendered instrumental music to extract prosody tokens as control conditions, denoted as $\bm{Q}_{p}$; (2) Let the policy model perform a completion conditioned on $\bm{Q}_{p}$ and $\bm{T}$ (where $\bm{T}$ can be any text), and denote the generated content-style tokens as $\hat{\bm{Q}}_{cs}$; (3) Using the decoder of the content-style tokenizer (Section~\ref{sec:tokenizer}), we reconstruct the chromagram features $\hat{\bm{x}}$ from $\hat{\bm{Q}}_{cs}$, and compute the cosine similarity between $\hat{\bm{x}}$ and the ground-truth chromagram features $\bm{x}$ of the singing voice ${u}$.\footnote{Note that in step (3), we could also use the flow-matching model and vocoder to perform inference and extract $\hat{\bm{x}}$ from the predicted waveform, but for training efficiency, we adopt the current approach.} 


\textbf{Multi-Objective Optimization}\quad
We apply the GRPO algorithm~\cite{grpo} to optimize the proposed multi-objective alignment task. Assuming the number of completions per prompt is $K$, for each prompt, after obtaining the intelligibility rewards ($\bm{r}_{int}$) and prosody similarity rewards ($\bm{r}_{pro}$) for all $K$ completions, we apply group normalization to both rewards separately. The sum of these normalized rewards serves as the final outcome supervision:
\begin{equation}
    A^{(i)} = \frac{r_{int}^{(i)} - \text{mean}(\bm{r}_{int})}{\text{std}(\bm{r}_{int})} + \frac{r_{pro}^{(i)} - \text{mean}(\bm{r}_{pro})}{\text{std}(\bm{r}_{pro})},
\label{eq:multi-reward-grpo}
\end{equation}
where $r_{int}^{(i)}$ and $r_{pro}^{(i)}$ denote the intelligibility reward and prosody similarity reward for the $i$-th completion, respectively, and $A^{(i)}$ represents the advantage.

\subsection{Inference-Time Controllability}\label{sec:inference-time-control}

\textbf{Text, Prosody, Style, and Timbre Control}\quad
After obtaining the pre-trained (or post-trained) models, we can perform inference following the pipeline illustrated in Figure~\ref{fig:vevo2-pipeline}, by utilizing different combinations of text, prosodic source, style reference, and timbre reference to achieve versatile synthesis, conversion, and editing tasks for both speech and singing voice.

During the content-style modeling stage, for example, when the prosodic source is not used, the inference paradigm is identical to classic zero-shot TTS~\cite{valle,seedtts,vevo,cosyvoice}. As illustrated in the TTS task in Figure~\ref{fig:vevo2-pipeline}, the process can be represented as [\colortwo{$\pmb{\bm{t}}$}, \colorone{$\pmb{\bm{t}}$}, \colortwo{$\pmb{\bm{cs}}$}] $\rightarrow$ [\colorpredict{$\pmb{\bm{\hat{cs}}}$}], where the style signals in \colorpredict{$\pmb{\bm{\hat{cs}}}$} are prompted by \colortwo{$\pmb{\bm{cs}}$} and generated through the in-context learning capabilities of the pre-trained model. Particularly, when both prosodic source and style reference are utilized as inputs, the inference paradigm takes the form [\colortwo{$\pmb{\bm{t}}$}, \colorone{$\pmb{\bm{t}}$}, \colortwo{$\pmb{\bm{p}}$}, \colorone{$\pmb{\bm{p}}$}, \colortwo{$\pmb{\bm{cs}}$}] $\rightarrow$ [\colorpredict{$\pmb{\bm{\hat{cs}}}$}]. In this context, for \colorpredict{$\pmb{\bm{\hat{cs}}}$}, we expect \colorone{$\pmb{\bm{p}}$} to provide \textit{coarse-grained} prosody contour signals, while \colortwo{$\pmb{\bm{cs}}$} contributes \textit{fine-grained} style details (e.g., the \textit{vibrato} patterns for singing). We will discuss this further through our experiments on style-converted singing voice conversion (Section~\ref{sec:expt-controllability-timbre-style}).

During the acoustic modeling stage, we can introduce timbre control by providing a timbre reference. Concretely, we concatenate the content-style tokens generated by the AR model with those extracted from the timbre reference, and jointly condition the FM model on the reference's Mel spectrogram, i.e., [\colortwo{$\pmb{\bm{cs}}$}, \colortwo{$\pmb{\bm{mel}}$}, \colorpredict{$\pmb{\bm{\hat{cs}}}$}] $\rightarrow$ [\colorpredict{$\pmb{\bm{\hat{mel}}}$}] of Figure~\ref{fig:vevo2-pipeline}. This dual conditioning guides the model to synthesize the target utterance with the linguistic, prosodic, and stylistic content from the AR output, while adopting the timbre characteristics embedded in the reference.
It is noteworthy that we can also use only the FM model to conduct style-preserved VC or SVC tasks. Specifically, the AR-generated content-style tokens can be replaced with those extracted directly from a source audio, i.e., [\colortwo{$\pmb{\bm{cs}}$}, \colortwo{$\pmb{\bm{mel}}$}, \colorone{$\pmb{\bm{cs}}$}] $\rightarrow$ [\colorpredict{$\pmb{\bm{\hat{mel}}}$}] of Figure~\ref{fig:vevo2-pipeline}.


\textbf{Duration Control}\quad
Some Non-AR based speech generation models have often featured the ability to control the total duration of generated speech~\cite{maskgct,f5tts}, while AR-based models are generally considered less capable of controlling output duration~\cite{audiolm,seedtts,cosyvoice,vevo,indextts2}. Interestingly, however, we discover that \vevo{} can effectively control the duration of generated results through the use of prosody tokens as input. This capability stems from our tokenizers' fixed frame rates (6.25 Hz and 12.5 Hz), which maintain a consistent 2:1 ratio between content-style and prosody token sequences during training. Consequently, during inference, we can control the duration of the generated output by applying simple linear scaling transformations to the chromagram features of the prosodic source, thereby modifying the length of prosody tokens. Such control mechanism is detailed in Section~\ref{sec:expt-controllability-duration}.

\textbf{Pitch Region Control}\quad
For voice conversion (VC) or singing voice conversion (SVC) tasks, due to potential significant differences in \textit{timbre types}~\cite{vocalset,singingvoice-science} between source and reference speakers, we often need to adjust the pitch region (i.e., apply \textit{pitch shift}) of the source waveform during inference to achieve higher speaker similarity~\cite{svcc-2023,amphion-svc,neucosvc}. In \vevo{}, we discover that pitch region control can also be achieved: by shifting the pitch of the source waveform before prosody token extraction, we can effectively control the pitch region of the generated speech or singing voice. Our experiments demonstrate that this approach is effective both when using only the FM stage and when combining AR and FM stages. Such control mechanism is detailed in Section~\ref{sec:expt-controllability-pitch}.

\section{Experimental Setup}\label{sec:expt}

\subsection{Data and Metrics}

\textbf{Training Data}\quad
For pre-training, we use Emilia~\cite{emilia} as the speech data (101K hours), and follow SingNet's pipeline~\cite{singnet} to prepare 7K hours of singing voice data (source-separated from in-the-wild songs from the Internet). 
For post-training, we utilize INTP~\cite{intp} (which contains 250K preference speech pairs) to train the intelligibility reward model. For the proposed multi-objective alignment task, we employ three types of prosody prompts: (1) speech data: 20K positive samples randomly selected from INTP, (2) singing voice data: 20K samples from M4Singer~\cite{m4singer}, and (3) instrumental sounds: rendered from MIDIs in M4Singer, covering 16 instruments across five categories: pianos, strings, woodwinds, brass, and folk instruments.


\textbf{Evaluation Data}\quad
We consider various settings to construct the evaluation set:
(1) For regular speech, we utilize SeedTTS test sets~\cite{seedtts}, which is widely adopted in the TTS field~\cite{maskgct,cosyvoice2,f5tts}. 
(2) For expressive speech, we employ Genshin-Voice~\cite{genshin_voice_dataset}, which consists of professional voice acting recordings from the game \textit{Genshin Impact}. 
(3) For singing voice, we use GTSinger~\cite{gtsinger} and SingStyle111~\cite{singstyle111}, which encompass a wide range of singing techniques, styles, and timbre types.
(4) Additionally, for two unique tasks: humming-to-singing and instrument-to-singing, we utilize humming~\cite{humtrans} and instrumental sound including flute~\cite{flute-dataset} and violin~\cite{violin-dataset} for the evaluation. 
All evaluations are conducted using target text in both English and Chinese languages.

\textbf{Evaluation Metrics}\quad
For the objective metrics, we evaluate the intelligibility (WER, $\downarrow$), speaker similarity (SIM, $\uparrow$), F0 correlation (FPC, $\uparrow$)~\cite{svcc-2023,amphion-svc,vevo}. For WER, we employ \texttt{Whisper-large-v3}~\cite{whisper} for English texts, and \texttt{Paraformer-zh}~\cite{paraformer,funasr} for Chinese texts. For SIM, we compute the cosine similarity between the WavLM TDNN~\cite{wavlm} speaker embeddings of generated samples and the reference speeches. 
For subjective metrics, we employ Comparative Mean Opinion Score (CMOS, rated from -2 to 2, $\uparrow$) to evaluate naturalness (N-CMOS) and similarity in speaker, prosody, melody, and style (SS-CMOS, PS-CMOS, MS-CMOS, and Style-CMOS). Besides, we use Melody-MOS to evaluate the melody-following ability (rated from 1 to 3, corresponding to ``unable to follow'', ``roughly following the melody contour'', and ``completely following all melody details''). Detailed backgrounds of subjects and definitions of all subjective metrics are provided in the supplementary materials.

\subsection{Implementation Details}\label{sec:impl-details}

\textbf{Prosody Tokenizer and Content-Style Tokenizer}\quad
This paper processes and trains on speech and singing voice data with a sampling rate of 24 kHz. For chromagram feature extraction, we employ Librosa's implementation\footnote{\url{https://librosa.org/doc/main/generated/librosa.feature.chroma_stft.html}} with the following parameters: number of FFT points set to 1920, hop length of 480, window size of 1920, and 24 chroma bins. For Whisper hidden features~\cite{whisper}, we utilize the encoder output from the pre-trained whisper-medium model\footnote{\url{https://huggingface.co/openai/whisper-medium}}. We adopt the VQ-VAE implementations from Amphion\footnote{\url{https://github.com/open-mmlab/Amphion/tree/main/models/vc/vevo}}~\cite{amphion,amphion_v0.2}, which serves for MaskGCT~\cite{maskgct} and Vevo~\cite{vevo}. Our only modification is the addition of a CNN-based downsampling layer before the encoder input and a CNN-based upsampling layer after the decoder output. The extracted chromagram features and Whisper features are both at 50 Hz. For the prosody tokenizer, the down(up)-sampling ratio is 8, resulting in a token rate of 6.25 Hz, with a codebook vocabulary size of 512. For the content-style tokenizer, the down(up)-sampling ratio is 4, yielding a token rate of 12.5 Hz, with a codebook vocabulary size of 16,384. The prosody tokenizer and content-style tokenizer contain 38M and 44M parameters, respectively. 

All models in this study use AdamW~\cite{adam,adamw} as the optimizer. For the two tokenizers, we use a peak learning rate of 1e-4, with linear warm-up for 10K steps followed by decay over the remaining training period. \textcolor{\revisioncolor}{Both tokenizers are trained on 4 NVIDIA A100 GPUs for 300K updates (about 48 hours).}

\textbf{Content-Style Modeling}\quad
During the content-style modeling stage, we initialize the AR model using Qwen2.5-0.5B\footnote{\url{https://huggingface.co/Qwen/Qwen2.5-0.5B}}~\cite{qwen2.5} and expand its original vocabulary to accommodate our proposed prosody tokens and content-style tokens. The total parameter count of the AR model is 509M. For the BPE-based text tokenizer, we adopted the same tokenization strategy as CosyVoice 2, which masks out one-to-many tokens to prevent excessive pronunciation length and mitigate corner cases caused by data sparsity~\cite{cosyvoice2}. During pre-training, we use a peak learning rate of 5e-4, with linear warm-up for 32K steps followed by decay over the remaining training period. \textcolor{\revisioncolor}{The model is pre-trained on 8 NVIDIA A100 GPUs for 500K updates (about 168 hours).}

\textbf{Acoustic Modeling}\quad
During the acoustic modeling stage, we adopted the flow-matching (FM) transformer architecture from Vevo~\cite{vevo}. The transformer consists of 16 layers with 16 attention heads and a hidden dimension of 1024. The FM model contains 363M parameters in total. To enhance training efficiency, we incorporated the REPA strategy~\cite{repa} and enforced alignment between the hidden output of the transformer's 5th layer and the W2v-BERT 2.0 features\footnote{\url{https://huggingface.co/facebook/w2v-bert-2.0}}~\cite{w2v-bert,maskgct} during the training process. We use a peak learning rate of 7.5e-5, with linear warm-up for 32K steps followed by decay over the remaining training period. \textcolor{\revisioncolor}{The FM model is trained on 8 NVIDIA A100 GPUs for 700K updates (about 168 hours).} To develop a unified vocoder for both speech and singing voice, we fine-tuned Vevo's speech vocoder\footnote{\url{https://huggingface.co/amphion/Vevo}}~\cite{vevo} using our training data (comprising 101K hours of speech and 7K hours of singing voice data). The vocoder, which is based on the Vocos architecture~\cite{vocos}, contains 255M parameters and underwent fine-tuning \textcolor{\revisioncolor}{on 4 NVIDIA A100 GPUs for 572K updates (about 72 hours)}.

\textbf{Post-training}\quad
During the post-training, for each training iteration with the 20K M4Singer samples, we randomly selected one instrument and rendered the corresponding audio using pretty\_midi\footnote{\url{https://github.com/craffel/pretty-midi}}. We use the original text of the sample and extract prosody tokens from these rendered instrumental sounds. We adopt the training framework of TRL\footnote{\url{https://huggingface.co/docs/trl/main/grpo_trainer}}. We use the learning rate of 5e-6, the batch size of 3, the rollout number of 8 and the KL coefficent is 0.1. \textcolor{\revisioncolor}{The model is post-trained on 8 NVIDIA A100 GPUs for 9K updates (about 48 hours).}

\section{Results and Analysis}\label{sec:results}

\begin{table*}[t]
\caption{Results on zero-shot TTS task. (\textbf{PT?}: Whether post-training is applied.)}
\vspace{-2mm}
\label{tab:expt-tts}
\begin{center}
\resizebox{\textwidth}{!}{
\begin{threeparttable}
    \begin{tabular}{l|c|c|c|c|rr|rr|rr|rr}
        \toprule
        \multicolumn{1}{c|}{\multirow{2}{*}{\makecell[c]{\\ \textbf{Model}} }} 
        & \multicolumn{1}{c|}{\multirow{2}{*}{\makecell[c]{\\ \textbf{\#Hz}}}}
        & \multicolumn{1}{c|}{\multirow{2}{*}{\makecell[c]{\\ \textbf{Data}\\ \textit{(\#hours)} }}}
        & \multicolumn{1}{c|}{\multirow{2}{*}{\makecell[c]{\\ \textcolor{\revisioncolor}{\textbf{\#Param}} }}}
        & \multicolumn{1}{c|}{\multirow{2}{*}{\makecell[c]{\\ \textbf{PT?} }}} 
        & \multicolumn{4}{c|}{\textbf{Expressive Speech}} 
        & \multicolumn{4}{c}{\textbf{Singing Voice}} \\
        \cmidrule(lr){6-9} \cmidrule(lr){10-13}
        & & & & & \textbf{WER} & \textbf{SIM} & \makecell[c]{\textbf{N-}\\ \textbf{CMOS}} & \makecell[c]{\textbf{SS-}\\ \textbf{CMOS}}
        & \textbf{WER} & \textbf{SIM} & \makecell[c]{\textbf{N-}\\ \textbf{CMOS}} & \makecell[c]{\textbf{SS-}\\ \textbf{CMOS}} \\
        \midrule
        \text{Ground Truth}
            & - & - & \textcolor{\revisioncolor}{-} & -
            & 10.91 & 0.687 & 1.47 $_{\scriptscriptstyle \pm \text{0.24}}$ & 0.87 $_{\scriptscriptstyle \pm \text{0.19}}$
            & 12.65 & 0.658 & 1.27 $_{\scriptscriptstyle \pm \text{0.14}}$ & 0.55 $_{\scriptscriptstyle \pm \text{0.16}}$ \\
        \midrule
        \text{F5-TTS}~\cite{f5tts}
            & - & \colorone{\text{101K}} & \textcolor{\revisioncolor}{336M} & \ding{55}
            & 11.77 & 0.695 & -1.05 $_{\scriptscriptstyle \pm \text{0.18}}$ & -0.06 $_{\scriptscriptstyle \pm \text{0.17}}$
            & 16.12 & 0.597 & -1.89 $_{\scriptscriptstyle \pm \text{0.07}}$ & -1.29 $_{\scriptscriptstyle \pm \text{0.10}}$ \\
        \text{MaskGCT}~\cite{maskgct}
            & \text{50} & \colorone{\text{101K}} & \textcolor{\revisioncolor}{1048M} & \ding{55}
            & 13.42 & \textbf{0.736} & -1.14 $_{\scriptscriptstyle \pm \text{0.21}}$ & \textbf{0.07} $_{\scriptscriptstyle \pm \text{0.31}}$
            & \underline{11.71} & \textbf{0.753} & -1.74 $_{\scriptscriptstyle \pm \text{0.14}}$ & -0.97 $_{\scriptscriptstyle \pm \text{0.13}}$ \\
        \text{CosyVoice 2}~\cite{cosyvoice}
            & \text{25} & \colorone{\text{167K}} & \textcolor{\revisioncolor}{581M} & \ding{51}
            & \textbf{11.20} & \underline{0.706} & \textbf{0.10} $_{\scriptscriptstyle \pm \text{0.29}}$ & -0.16 $_{\scriptscriptstyle \pm \text{0.13}}$
            & 16.18 & 0.659 & -1.68 $_{\scriptscriptstyle \pm \text{0.15}}$ & -1.12 $_{\scriptscriptstyle \pm \text{0.16}}$ \\
        \midrule
        \multirow{3}{*}{\text{\vevo{}-base}}
            & \multirow{3}{*}{\makecell[c]{\text{12.5}}} & \colorone{\text{101K}} & \multirow{3}{*}{\makecell[c]{\textcolor{\revisioncolor}{872M}}} & \ding{55}
            & 15.52 & 0.677 & -0.84 $_{\scriptscriptstyle \pm \text{0.15}}$ & -0.06 $_{\scriptscriptstyle \pm \text{0.27}}$
            & 19.39 & 0.658 & -1.02 $_{\scriptscriptstyle \pm \text{0.16}}$ & -0.48 $_{\scriptscriptstyle \pm \text{0.18}}$ \\
        
            & & \colortwo{\text{7K}} & & \ding{55}
            & 37.57 & 0.611 & -1.25 $_{\scriptscriptstyle \pm \text{0.26}}$ & -0.57 $_{\scriptscriptstyle \pm \text{0.17}}$
            & 25.38 & 0.692 & -0.69 $_{\scriptscriptstyle \pm \text{0.26}}$ & -0.36 $_{\scriptscriptstyle \pm \text{0.17}}$ \\
            & & \colorone{\text{101K}}, \colortwo{\text{7K}} & & \ding{55}
            & 14.32 & 0.681 & -0.49 $_{\scriptscriptstyle \pm \text{0.24}}$ & -0.03 $_{\scriptscriptstyle \pm \text{0.19}}$
            & 15.78 & 0.708 & \underline{-0.28} $_{\scriptscriptstyle \pm \text{0.14}}$ & \underline{-0.09} $_{\scriptscriptstyle \pm \text{0.21}}$ \\
            \midrule
            \vevo{} & 12.5 & \colorone{\text{101K}}, \colortwo{\text{7K}} & \textcolor{\revisioncolor}{872M} & \ding{51}
            & \underline{11.48} & 0.689 & \underline{0.00} \textcolor{white}{$_{\scriptscriptstyle \pm \text{0.00}}$} & \underline{0.00} \textcolor{white}{$_{\scriptscriptstyle \pm \text{0.00}}$}
             & \textbf{7.66} & \underline{0.725} & \textbf{0.00} \textcolor{white}{$_{\scriptscriptstyle \pm \text{0.00}}$} & \textbf{0.00} \textcolor{white}{$_{\scriptscriptstyle \pm \text{0.00}}$} \\
        \bottomrule
    \end{tabular}
    \begin{tablenotes}
        \footnotesize{
            \item[*] \textbf{\#Hz}: Frame rate of the content-style tokenizer. \textbf{Data}: Pre-training data, where \colorone{\textbf{blue}} and \colortwo{\textbf{red}} denote \colorone{\textbf{speech}} and \colortwo{\textbf{singing}} data. \textcolor{\revisioncolor}{\textbf{\#Param}: The number of learnable parameters.} The results on the regular speech domain are provided in the supplementary materials. The best and the second best result (excluding ground truth) are shown in \textbf{bold} and by \underline{underlined}.
        }
        \vspace{-4mm}
    \end{tablenotes}
\end{threeparttable}
}
\end{center}
\end{table*}

\subsection{Benefits of Unified Modeling for Speech and Singing Voice}\label{sec:expt-mutual-benefits}

To investigate whether speech and singing voice generation can be mutually beneficial through unified modeling, we examine the impact of different pre-training data distributions on \vevo{}. Specifically, we conduct experiments on zero-shot TTS tasks and evaluate the performance using reference prompts from different distributions. For comparison, we select models with varying content-style tokenizer frame rates: MaskGCT~\cite{maskgct} (50 Hz), CosyVoice 2~\cite{cosyvoice2} (25 Hz), and F5-TTS~\cite{f5tts} (a single-stage model without content-style tokenizer). The results are presented in Table~\ref{tab:expt-tts}, which demonstrates the following findings: 

\textbf{(1) Speech and singing voice generation mutually benefit from unified pre-training:} Comparing \vevo{} models with different pre-training data, we observe that on expressive speech domain, jointly using speech and singing data for pre-training improves all metrics (WER: 15.52 $\rightarrow$ 14.32, SIM: 0.677 $\rightarrow$ 0.681, N-CMOS: -0.84 $\rightarrow$ -0.49, SS-CMOS: -0.06 $\rightarrow$ -0.03). On singing voice domain, we observe that using only singing voice data for pre-training yields lower intelligibility compared to using speech data alone (WER: 25.38 vs. 19.39), likely due to the smaller size of the singing voice corpus. Furthermore, when combining speech and singing voice data for pre-training, \vevo{} demonstrates superior intelligibility (WER: 19.39 $\rightarrow$ 15.78) and improved naturalness and similarity (N-CMOS: -1.02 $\rightarrow$ -0.28, SS-CMOS: -0.48 $\rightarrow$ -0.09). 

\textbf{(2) Effectiveness of the proposed post-training:} The proposed post-training brings further overall improvements to the pre-trained model (\vevo{}-base), including enhanced intelligibility (WER), speaker similarity (SIM and SS-CMOS), and naturalness (N-CMOS). This demonstrates the effectiveness and significant potential of post-training techniques in controllable speech and singing voice generation. 

\textbf{(3) Competitive performance of \vevo{} compared to existing models:} Compared to the baselines, the post-trained \vevo{} matches and even outperforms them on certain metrics in the expressive speech domain, despite its lower frame rate in the content-style tokenizer. More significantly, on singing voice domain, \vevo{} substantially outperforms existing zero-shot TTS models in subjective evaluations (N-CMOS: gaps greater than 1.5; SS-CMOS: gaps greater than 0.9), benefiting greatly from the incorporation of singing voice data during training. This validates \vevo{}'s effectiveness as a unified controllable speech and singing voice generation model.


\subsection{Controllability over Text and Prosody}\label{sec:expt-controllability-text-prosody}

\begin{table*}[t]
\caption{Results on zero-shot SVS, speech editing, and singing lyric editing tasks.}
\vspace{1mm}
\label{tab:expt-svs-editing}
\centering
\begin{subtable}{0.95\textwidth}
    \resizebox{\textwidth}{!}{%
    \begin{tabular}{l|c|rrr|rc|rrr|rc}
        \toprule
        \midrule
        \multicolumn{12}{c}{\textbf{\textit{Singing Voice Synthesis}}} \\
        \midrule \midrule
        \multicolumn{1}{c|}{\multirow{3}{*}{\textbf{Model}}}
        & \multicolumn{1}{c|}{\multirow{3}{*}{\textcolor{\revisioncolor}{\textbf{\#Param}} }}
        & \multicolumn{5}{c|}{\textbf{English}} 
        & \multicolumn{5}{c}{\textbf{Chinese}} \\
        \cmidrule(lr){3-7} \cmidrule(lr){8-12}
        & & \textbf{WER} & \textbf{SIM} & \textbf{FPC} & \makecell[c]{\textbf{N-}\\ \textbf{CMOS}}  & \makecell[c]{\textbf{Melody-}\\ \textbf{MOS}}
        & \textbf{WER} & \textbf{SIM} & \textbf{FPC} & \makecell[c]{\textbf{N-}\\ \textbf{CMOS}} & \makecell[c]{\textbf{Melody-}\\ \textbf{MOS}}\\
        \midrule
        \text{Ground Truth}
            & \textcolor{\revisioncolor}{-} & 17.89 & 0.640 & 0.815 & - \textcolor{white}{$_{\scriptscriptstyle \pm \text{0.00}}$} & - \textcolor{white}{$_{\scriptscriptstyle \pm \text{0.00}}$}
            & 7.40 & 0.675 & 0.782 & - \textcolor{white}{$_{\scriptscriptstyle \pm \text{0.00}}$} & - \textcolor{white}{$_{\scriptscriptstyle \pm \text{0.00}}$}  \\
        \midrule
        \text{StyleSinger}~\cite{stylesinger}
            & \textcolor{\revisioncolor}{42M} & - & - & - & - \textcolor{white}{$_{\scriptscriptstyle \pm \text{0.00}}$} & - \textcolor{white}{$_{\scriptscriptstyle \pm \text{0.00}}$}
            & 14.04 & 0.555 & 0.670 & -0.37 $_{\scriptscriptstyle \pm \text{0.16}}$  & 2.33 $_{\scriptscriptstyle \pm \text{0.36}}$ \\
        \text{TCSinger}~\cite{tcsinger}
            & \textcolor{\revisioncolor}{330M} & 40.83 & 0.605 & 0.724 & -0.09 $_{\scriptscriptstyle \pm \text{0.12}}$  & \textbf{2.55} $_{\scriptscriptstyle \pm \text{0.27}}$
            & 14.26 & 0.651 & 0.707 & -0.28 $_{\scriptscriptstyle \pm \text{0.17}}$  & \textbf{2.58} $_{\scriptscriptstyle \pm \text{0.19}}$  \\
        \midrule
        \text{\vevo{}-base}
            & \textcolor{\revisioncolor}{872M} & 60.83 & 0.586 & 0.719 & - \textcolor{white}{$_{\scriptscriptstyle \pm \text{0.00}}$} & - \textcolor{white}{$_{\scriptscriptstyle \pm \text{0.00}}$}
            & 23.60 & 0.653 & 0.674 & - \textcolor{white}{$_{\scriptscriptstyle \pm \text{0.00}}$} & - \textcolor{white}{$_{\scriptscriptstyle \pm \text{0.00}}$} \\
        \text{\vevo{}}
            & \textcolor{\revisioncolor}{872M} & \textbf{24.77} & \textbf{0.643} & \textbf{0.752} & \textbf{0.00} \textcolor{white}{$_{\scriptscriptstyle \pm \text{0.00}}$}  & 2.29 $_{\scriptscriptstyle \pm \text{0.26}}$
            & \textbf{9.83} & \textbf{0.669} & \textbf{0.710} & \textbf{0.00} \textcolor{white}{$_{\scriptscriptstyle \pm \text{0.00}}$}  & 2.44 $_{\scriptscriptstyle \pm \text{0.18}}$ \\
    \end{tabular}
    }
\end{subtable}
\hfill
\begin{subtable}{0.95\textwidth}
    \resizebox{\textwidth}{!}{%

    \begin{threeparttable}
        \begin{tabular}{l|c|rrr|rr|rrr|rr}
        \midrule
        \midrule
        \multicolumn{11}{c}{\textbf{\textit{Editing}}} \\
        \midrule \midrule
        \multicolumn{1}{c|}{\multirow{3}{*}{\textbf{Model}}}
        & \multicolumn{1}{c|}{\multirow{3}{*}{\textcolor{\revisioncolor}{\textbf{\#Param}}}} & \multicolumn{5}{c|}{\textbf{Expressive Speech}} 
        & \multicolumn{5}{c}{\textbf{Singing Voice}} \\
        \cmidrule(lr){3-7} \cmidrule(lr){8-12}
        & & \textbf{WER} & \textbf{SIM} & \textbf{FPC} & \makecell[c]{\textbf{N-}\\ \textbf{CMOS}} & \makecell[c]{\textbf{PS-}\\ \textbf{CMOS}}
        & \textbf{WER} & \textbf{SIM} & \textbf{FPC} & \makecell[c]{\textbf{N-}\\ \textbf{CMOS}} & \makecell[c]{\textbf{MS-}\\ \textbf{CMOS}} \\
        \midrule
        \text{SSR-Speech}~\cite{ssrspeech}
            & \textcolor{\revisioncolor}{830M} & 29.56 & 0.588 & 0.721 & -1.36 $_{\scriptscriptstyle \pm \text{0.09}}$ & -0.89 $_{\scriptscriptstyle \pm \text{0.14}}$
            & 51.92 & 0.710 & 0.804 & -1.43 $_{\scriptscriptstyle \pm \text{0.17}}$ & -1.10 $_{\scriptscriptstyle \pm \text{0.08}}$ \\
        \text{F5-TTS}~\cite{f5tts}
            & \textcolor{\revisioncolor}{336M} & 21.35 & 0.733 & 0.730 & -0.19 $_{\scriptscriptstyle \pm \text{0.12}}$ & -0.21 $_{\scriptscriptstyle \pm \text{0.16}}$  
            & 29.78 & 0.784 & 0.821 & -1.15 $_{\scriptscriptstyle \pm \text{0.23}}$  & -0.97 $_{\scriptscriptstyle \pm \text{0.19}}$  \\
        \midrule
        \text{\vevo{}-base}
            & \textcolor{\revisioncolor}{872M} & 23.54 & 0.795 & 0.782 & - \textcolor{white}{$_{\scriptscriptstyle \pm \text{0.00}}$}  & - \textcolor{white}{$_{\scriptscriptstyle \pm \text{0.00}}$} 
            & 29.69 & 0.841 & 0.872 & - \textcolor{white}{$_{\scriptscriptstyle \pm \text{0.00}}$} & - \textcolor{white}{$_{\scriptscriptstyle \pm \text{0.00}}$}  \\
        \text{\vevo{}}
            & \textcolor{\revisioncolor}{872M} & \textbf{16.83} & \textbf{0.799} & \textbf{0.792} & \textbf{0.00} \textcolor{white}{$_{\scriptscriptstyle \pm \text{0.00}}$} & \textbf{0.00} \textcolor{white}{$_{\scriptscriptstyle \pm \text{0.00}}$}
            & \textbf{17.98} & \textbf{0.848} & \textbf{0.877} & \textbf{0.00} \textcolor{white}{$_{\scriptscriptstyle \pm \text{0.00}}$} & \textbf{0.00} \textcolor{white}{$_{\scriptscriptstyle \pm \text{0.00}}$} \\
        \bottomrule
    \end{tabular}
        \begin{tablenotes}
        \footnotesize{
            \item[*] \vevo{}-base here is pre-trained on both speech and singing data, corresponding to the row ``\vevo{}-base (\colorone{101K}, \colortwo{7K})" in Table~\ref{tab:expt-tts}.
        }
        \end{tablenotes}
    \end{threeparttable}
    }
\end{subtable}
\end{table*}

\begin{table*}[t]
\caption{Results on zero-shot SVC task. \vevo{}-FM denotes the flow-matching model of \vevo{}. }
\vspace{-1mm}
\label{tab:expt-svc}
\begin{center}
\resizebox{\textwidth}{!}{
\begin{threeparttable}
    \begin{tabular}{l|c|rr|rrr|rr|rrr}
        \toprule
        \multicolumn{1}{c|}{\multirow{3}{*}{\textbf{Model}}}
        & \multicolumn{1}{c|}{\multirow{3}{*}{\textcolor{\revisioncolor}{\textbf{\#Param}}}} & \multicolumn{5}{c|}{\textbf{English}} 
        & \multicolumn{5}{c}{\textbf{Chinese}} \\
        \cmidrule(lr){3-7} \cmidrule(lr){8-12}
        & & \textbf{WER} & \textbf{SIM} & \makecell[c]{\textbf{N-}\\ \textbf{CMOS}}  & \makecell[c]{\textbf{Style-}\\ \textbf{CMOS}} & \makecell[c]{\textbf{Melody-}\\ \textbf{MOS}}
        & \textbf{WER} & \textbf{SIM} & \makecell[c]{\textbf{N-}\\ \textbf{CMOS}}  & \makecell[c]{\textbf{Style-}\\ \textbf{CMOS}} & \makecell[c]{\textbf{Melody-}\\ \textbf{MOS}} \\
        \midrule
        \text{Ground Truth}
            & \textcolor{\revisioncolor}{-} & 16.48 & - & - \textcolor{white}{$_{\scriptscriptstyle \pm \text{0.00}}$} & - \textcolor{white}{$_{\scriptscriptstyle \pm \text{0.00}}$} & - \textcolor{white}{$_{\scriptscriptstyle \pm \text{0.00}}$}
            & 12.82 & - & - \textcolor{white}{$_{\scriptscriptstyle \pm \text{0.00}}$} & - \textcolor{white}{$_{\scriptscriptstyle \pm \text{0.00}}$} & - \textcolor{white}{$_{\scriptscriptstyle \pm \text{0.00}}$} \\
        \midrule
        \text{FACodec}~\cite{naturalspeech3}
            & \textcolor{\revisioncolor}{139M} & 32.81 & 0.434 & - \textcolor{white}{$_{\scriptscriptstyle \pm \text{0.00}}$} & - \textcolor{white}{$_{\scriptscriptstyle \pm \text{0.00}}$} & - \textcolor{white}{$_{\scriptscriptstyle \pm \text{0.00}}$}
            & 36.97 & 0.473 & - \textcolor{white}{$_{\scriptscriptstyle \pm \text{0.00}}$} & - \textcolor{white}{$_{\scriptscriptstyle \pm \text{0.00}}$} & - \textcolor{white}{$_{\scriptscriptstyle \pm \text{0.00}}$} \\
        \text{Vevo-FM}~\cite{vevo}
            & \textcolor{\revisioncolor}{334M} & 24.05 & 0.567 & -0.64 $_{\scriptscriptstyle \pm \text{0.07}}$ & -0.45 $_{\scriptscriptstyle \pm \text{0.16}}$  & 0.71 $_{\scriptscriptstyle \pm \text{0.21}}$
            & 22.85 & 0.610 & -0.55 $_{\scriptscriptstyle \pm \text{0.19}}$ & -0.44 $_{\scriptscriptstyle \pm \text{0.08}}$ & 0.88 $_{\scriptscriptstyle \pm \text{0.27}}$ \\
        \text{CosyVoice2-FM}~\cite{cosyvoice2}
            & \textcolor{\revisioncolor}{75M} & 23.59 & 0.553 & -0.64 $_{\scriptscriptstyle \pm \text{0.11}}$ & -0.63 $_{\scriptscriptstyle \pm \text{0.18}}$  & 2.25 $_{\scriptscriptstyle \pm \text{0.15}}$
            & 20.14 & 0.589 & -0.63 $_{\scriptscriptstyle \pm \text{0.15}}$ & -0.54 $_{\scriptscriptstyle \pm \text{0.07}}$  & 1.65 $_{\scriptscriptstyle \pm \text{0.23}}$ \\
        \midrule
        \text{NeuCoSVC 2}~\cite{neucosvc}
            & \textcolor{\revisioncolor}{14M} & 30.61 & 0.481 & - \textcolor{white}{$_{\scriptscriptstyle \pm \text{0.00}}$} & - \textcolor{white}{$_{\scriptscriptstyle \pm \text{0.00}}$} & - \textcolor{white}{$_{\scriptscriptstyle \pm \text{0.00}}$}
            & 32.26 & 0.519 & - \textcolor{white}{$_{\scriptscriptstyle \pm \text{0.00}}$} & - \textcolor{white}{$_{\scriptscriptstyle \pm \text{0.00}}$} & - \textcolor{white}{$_{\scriptscriptstyle \pm \text{0.00}}$} \\
        \text{SeedVC (SVC)}~\cite{seedvc}
            & \textcolor{\revisioncolor}{200M} & \underline{22.86} & 0.508 & \underline{-0.05} $_{\scriptscriptstyle \pm \text{0.13}}$ & -0.24 $_{\scriptscriptstyle \pm \text{0.15}}$ & {2.89} $_{\scriptscriptstyle \pm \text{0.09}}$
            & \underline{15.65} & 0.550 & -0.33 $_{\scriptscriptstyle \pm \text{0.21}}$ & -0.54 $_{\scriptscriptstyle \pm \text{0.14}}$  & \textbf{2.93} $_{\scriptscriptstyle \pm \text{0.14}}$  \\
        \midrule
        \text{\vevo{}-FM}
            & \textcolor{\revisioncolor}{363M} & 29.82 & \underline{0.587} & -0.12 $_{\scriptscriptstyle \pm \text{0.16}}$ & \underline{-0.11} $_{\scriptscriptstyle \pm \text{0.12}}$  & \textbf{2.91} $_{\scriptscriptstyle \pm \text{0.12}}$
            & 22.54 & \underline{0.611} & \underline{-0.03} $_{\scriptscriptstyle \pm \text{0.30}}$ & \underline{-0.32} $_{\scriptscriptstyle \pm \text{0.17}}$ & \underline{2.90} $_{\scriptscriptstyle \pm \text{0.08}}$ \\
        \text{\vevo{}}
            & \textcolor{\revisioncolor}{872M} & \textbf{11.64} & \textbf{0.601} & \textbf{0.00} \textcolor{white}{$_{\scriptscriptstyle \pm \text{0.00}}$} &  \textbf{0.00} \textcolor{white}{$_{\scriptscriptstyle \pm \text{0.00}}$} & 2.24 $_{\scriptscriptstyle \pm \text{0.18}}$
            & \textbf{14.53} & \textbf{0.623} & \textbf{0.00} \textcolor{white}{$_{\scriptscriptstyle \pm \text{0.00}}$} & \textbf{0.00} \textcolor{white}{$_{\scriptscriptstyle \pm \text{0.00}}$}  & 2.38 $_{\scriptscriptstyle \pm \text{0.25}}$  \\
        \bottomrule
    \end{tabular}
    \begin{tablenotes}
        \footnotesize{
            \item[*] The best and the second best result (excluding ground truth) is shown in \textbf{bold} and by \underline{underlined}.
        }
    \end{tablenotes}
\end{threeparttable}
}
\end{center}
\end{table*}

To evaluate \vevo{}'s controllability with respect to text and prosody (melody), we conduct various tasks including zero-shot SVS, speech editing, singing lyric editing, humming-to-singing, and instrument-to-singing (the latter two tasks will be discussed in Section~\ref{sec:ablation-grpo}). 

\textbf{Zero-Shot SVS}\quad
Given a target text, a music notation like MIDI, and a reference singer's waveform, a zero-shot SVS model is required to generate singing voice that matches the content of the target text and music notation while preserving the timbre and style of the reference singer~\cite{stylesinger,tcsinger,spsinger}. For \vevo{}, we render MIDI into piano waveform and use it as the prosodic source, which can also be considered as instrument-to-singing. 
The results are shown in Table~\ref{tab:expt-svs-editing}, where we include StyleSinger~\cite{stylesinger} and TCSinger~\cite{tcsinger} as two baselines. 

It demonstrates that: (1) After post-training, \vevo{} shows improved text-following ability and speaker similarity in both English and Chinese. We suspect that this is because that during pre-training, \vevo{} had not encountered prosody tokens extracted from instrumental sounds, resulting in unstable performance. By post-training, we not only supplement with MIDI-rendered instrumental sound data, but also adopt the specific alignment learning, effectively eliciting the model's potential. (2) Compared to the baselines, the post-trained \vevo{} demonstrates superior text-following ability (WER) and naturalness (N-CMOS). Regarding melody-following, although the post-trained \vevo{}'s Melody-MOS is slightly lower than TCSinger's, its scores for both English and Chinese exceed 2.0, indicating that \vevo{} is at least capable of ``roughly following the melody contour'' (as defined by Melody-CMOS score of 2.0) in most cases.

\textbf{Zero-Shot Editing}\quad
In addition to SVS, we adopt editing tasks to evaluate \vevo{}'s text and prosody controllability. The used evaluation sets are constructed with DeepSeek-V3~\cite{deepseek-v3}. Specifically, we utilized the same expressive speech and singing voice evaluation data from our zero-shot TTS tasks. For each sample's text, we generated edited versions using DeepSeek-V3 with the following prompt: \textit{"Please edit the following text by adding, deleting, or modifying content as needed. Note: (1) Ignore any existing punctuation errors in the original text and ensure proper punctuation in the edited version. (2) The edited version must differ from the original text. (3) Return only the final edited text without any additional explanations."} 

\vevo{}'s advantage in editing tasks stems from its use of prosody tokens as explicit control, enabling it to preserve the prosody (melody) of the raw voice during editing. For instance, in singing lyric editing tasks~\cite{editsinger,unisinger}, \vevo{} can modify only the lyrics while maintaining the original singing voice melody. The results are shown in Table~\ref{tab:expt-svs-editing}, where we include two speech editing baselines, SSR-Speech~\cite{ssrspeech} and F5-TTS~\cite{f5tts}. It can be observed that compared to the baselines, the post-trained \vevo{} not only achieves better intelligibility (WER), naturalness (N-CMOS), and speaker similarity (SIM), but also demonstrates superior prosody preservation of the original voice (PS-CMOS), particularly in the singing voice domain (MS-CMOS). We will provide more audio samples at our website to further demonstrate \vevo{}'s advantages in singing lyric editing tasks.

\begin{table*}[t]
\begin{minipage}{0.73\textwidth}
    \caption{Results on zero-shot accent and emotion conversion tasks.}
    \vspace{-1mm}
    \label{tab:expt-accent-emotion-conversion}
    \begin{center}
    \resizebox{\textwidth}{!}{
    \begin{threeparttable}
        \begin{tabular}{l|c|rccc|rccc}
            \toprule
            \multicolumn{1}{c|}{\multirow{2}{*}{\textbf{Model}}}
            & \multicolumn{1}{c|}{\multirow{2}{*}{\textcolor{\revisioncolor}{\textbf{\#Param}}}} & \multicolumn{4}{c|}{\textbf{\textit{Accent Conversion}}} 
            & \multicolumn{4}{c}{\textbf{\textit{Emotion Conversion}}} \\
            \cmidrule(lr){3-6} \cmidrule(lr){7-10}
            & & \textbf{WER} & \textbf{SIM} & \textbf{A-SIM} & \textbf{UTMOS}
            & \textbf{WER} & \textbf{SIM} & \textbf{E-SIM} & \textbf{UTMOS} \\
            \midrule
            Vevo~\cite{vevo} & \textcolor{\revisioncolor}{797M}
                & 30.37 & 0.702 & 0.526 & 3.49
                & 13.31 & 0.697 & 0.406 & 3.95 \\
            \midrule
            \vevo{}-base & \textcolor{\revisioncolor}{872M}
                & 10.08 & 0.760 & 0.537 & 3.72
                & 11.36 & 0.700 & 0.400 & 3.82 \\
            \vevo{} & \textcolor{\revisioncolor}{872M}
                & 7.73 & 0.768 & 0.529 & 3.76
                & 7.89 & 0.714 & 0.402 & 3.89 \\
            \bottomrule
        \end{tabular}
        \begin{tablenotes}
            \footnotesize{
                \item[*] \textbf{A-SIM}: Accent-SIM. \textbf{E-SIM}: Emotion-SIM.
            }
        \end{tablenotes}
    \end{threeparttable}
    }
    \end{center}
\end{minipage}
\hfill
\begin{minipage}{0.25\textwidth}
    \caption{Effect of duration control of \vevo{}.}
    \label{tab:expt-duration-control}
    \begin{center}
    \resizebox{\textwidth}{!}{
    \begin{threeparttable}
        \begin{tabular}{l|c|c}
            \toprule
            \textbf{Task} & \textbf{DDUR} ($\downarrow$) & \makecell[c]{\textbf{DC} ($\uparrow$)} \\
            \midrule
            \textbf{\text{SE}} & 0.22s & 97.1\% \\
            \textbf{\text{SLE}} & 0.20s & 98.4\% \\
            \textbf{\text{SVC}} & 0.12s & 98.4\% \\
            \bottomrule
        \end{tabular}
        \begin{tablenotes}
            \footnotesize{
                \item[*] \textbf{SE}: Speech Editing. \textbf{SLE}: Singing Lyric Editing.
            }
        \end{tablenotes}
        \vspace{-4mm}
    \end{threeparttable}
    }
    \end{center}
\end{minipage}
\end{table*}

\subsection{Controllability over Timbre and Style}\label{sec:expt-controllability-timbre-style}
To evaluate \vevo{}'s controllability of timbre and style, we conduct tasks including zero-shot VC (see the supplementary materials), SVC, and speech style conversion (for accent and emotion).

\textbf{Zero-Shot SVC}\quad
For zero-shot SVC, we include five baselines: three models pre-trained only on speech data (FACodec~\cite{naturalspeech3}, Vevo~\cite{vevo}, and CosyVoice 2~\cite{cosyvoice2}), and two models pre-trained on both speech and singing voice data (NeuCoSVC 2~\cite{neucosvc} and SeedVC (SVC)~\cite{seedvc}). Notably, for Vevo and CosyVoice 2, we utilize their flow-matching (FM) models. Similarly, we include \vevo{}'s FM model as a baseline (denoted as \vevo{}-FM). The results in Table~\ref{tab:expt-svc} reveal two key findings:

(1) Comparing \vevo{}-FM with the existing baselines, \vevo{}-FM demonstrates superior timbre-disentanglement (higher SIM for both English and Chinese), validating its effectiveness for timbre control. Additionally, it exhibits strong melody modeling capability (high Melody-MOS), attributable to the explicit modeling of chromagram features in our content-style tokenizer. However, due to the lower frame rate of content-style tokens (12.5 Hz), \vevo{}-FM shows slightly lower intelligibility compared to Vevo-FM (50 Hz), CosyVoice2-FM (25 Hz), and SeedVC (using continuous features).

(2) Comparing \vevo{}-FM and \vevo{}, there are two interesting observations:
    \textit{a)} \vevo{} shows significant improvements in intelligibility, which we attribute to both the explicit use of text input and intelligibility alignment during post-training.
    \textit{b)} Although \vevo{}'s Melody-MOS is lower than \vevo{}-FM's, its absolute value still exceeds 2, indicating its ability to ``roughly follow the melody contour''. This trade-off benefits style-converted SVC\footnote{For style-converted SVC, we aim for the generated singing voice to follow the source's \textit{coarse-grained melody contour} while imitating \textit{fine-grained expression details} (such as \textit{vibrato}) from the \textit{reference}~\cite{vevo,svcc2025}.}: specifically, \vevo{} achieves better style imitation performance (higher Style-CMOS), leading to improved speaker similarity (higher SIM). Similar characteristics are also observed in style-converted VC tasks~\cite{vevo}.

\textbf{Zero-Shot Accent and Emotion Conversion}\quad
Table~\ref{tab:expt-accent-emotion-conversion} demonstrates \vevo{}'s performance on zero-shot accent and emotion conversion tasks. We benchmarked against Vevo~\cite{vevo} and adopted their evaluation sets and two style conversion metrics: Accent-SIM (computed using a pre-trained accent classifier\footnote{\url{https://huggingface.co/Jzuluaga/accent-id-commonaccent_ecapa}}) and Emotion-SIM (calculated via a pre-trained emotion classifier\footnote{\url{https://github.com/ddlBoJack/emotion2vec}}) to assess style conversion performance. The results indicate that \vevo{} achieves comparable style similarity scores to Vevo while exhibiting superior performance in both intelligibility and speaker similarity metrics.

\subsection{Controllability over Duration}\label{sec:expt-controllability-duration}
As discussed in Section~\ref{sec:inference-time-control}, we discovered that \vevo{} enables effective duration control of generated outputs through the utilization of prosody tokens as input. This capability arises from a fundamental characteristic: both our proposed prosody tokenizer and content-style tokenizer operate at fixed frame rates (6.25 Hz and 12.5 Hz, respectively). During training, this results in a consistent linear relationship between the sequence lengths of content-style tokens and prosody tokens, specifically maintaining a 2:1 ratio. The AR model readily learns this straightforward length pattern. Consequently, during inference, we can control the output duration by applying linear scaling transformations to the chromagram features of the prosodic source—leveraging the inherent robustness of chromagram features to such transformations—thereby modulating the length of prosody tokens.


In our experiments, we investigated \vevo{}'s duration control effectiveness across three tasks: speech editing (SE), singing lyric editing (SLE), and style-converted SVC. We adopt the metric, \textbf{Differences of Duration (DDUR, $\downarrow$)}~\cite{non-parallel-seq2seq-vc,vevo}, which measures the average absolute difference in seconds between the actual generated duration ($\hat{d}$) and the target duration ($d$). Additionally, we introduce \textbf{Duration Consistency (DC, $\uparrow$)} to evaluate the relative accuracy of duration control:
\begin{equation}
     \textbf{DDUR} = \text{abs}(\hat{d} - {d}),\quad \textbf{DC} = \frac{\text{abs}(\hat{d} - {d})}{d}.
\end{equation}

For editing tasks, following F5-TTS~\cite{f5tts}, we calculate the target duration based on the linear relationship between the lengths of edited and raw text. For the style-converted SVC, we constrain the target duration to match the source waveform duration. As demonstrated in Table~\ref{tab:expt-duration-control}, \vevo{} achieved over 97\% duration consistency across all three tasks, exhibiting good duration controllability despite its autoregressive nature.

\begin{figure*}[t]
\begin{minipage}{0.56\textwidth}
\begin{center}
\resizebox{\textwidth}{!}{
\begin{threeparttable}
    \begin{tabular}{l|ccc|ccc}
        \toprule
        \multicolumn{1}{c|}{\multirow{2}{*}{\textbf{Model}}}
        & \multicolumn{3}{c|}{\textbf{\textit{Humming-to-Singing}}} 
        & \multicolumn{3}{c}{\textbf{\textit{Instrument-to-Singing}}} \\
        \cmidrule(lr){2-4} \cmidrule(lr){5-7}
        & \textbf{WER} & \textbf{SIM} & \textbf{FPC}
        & \textbf{WER} & \textbf{SIM} & \textbf{FPC} \\
        \midrule
        Base
            & 32.86 & 0.534 & 0.769
            & 40.38 & 0.503 & 0.716 \\
        \ \ \ \     \textit{w/ Intell}
            & 17.49 & 0.581 & 0.774
            & 20.03 & 0.502 & 0.731 \\
        \ \ \ \ \ \ \ \ \textit{w/ Prosody}
            & 17.27 & 0.585 & 0.784 
            & 17.94 & 0.525 & 0.745 \\
        \bottomrule
    \end{tabular}
    \begin{tablenotes}
        \footnotesize{
            \item[*] The first and last model represent \vevo{}-base and \vevo{}, respectively. \textit{w/ Intell}: Only intelligibility reward is used for post-training.
        }
    \end{tablenotes}
\end{threeparttable}
}
\end{center}
\end{minipage}
\hfill
\begin{minipage}{0.42\textwidth}
    \centering
    \includegraphics[width=\textwidth]{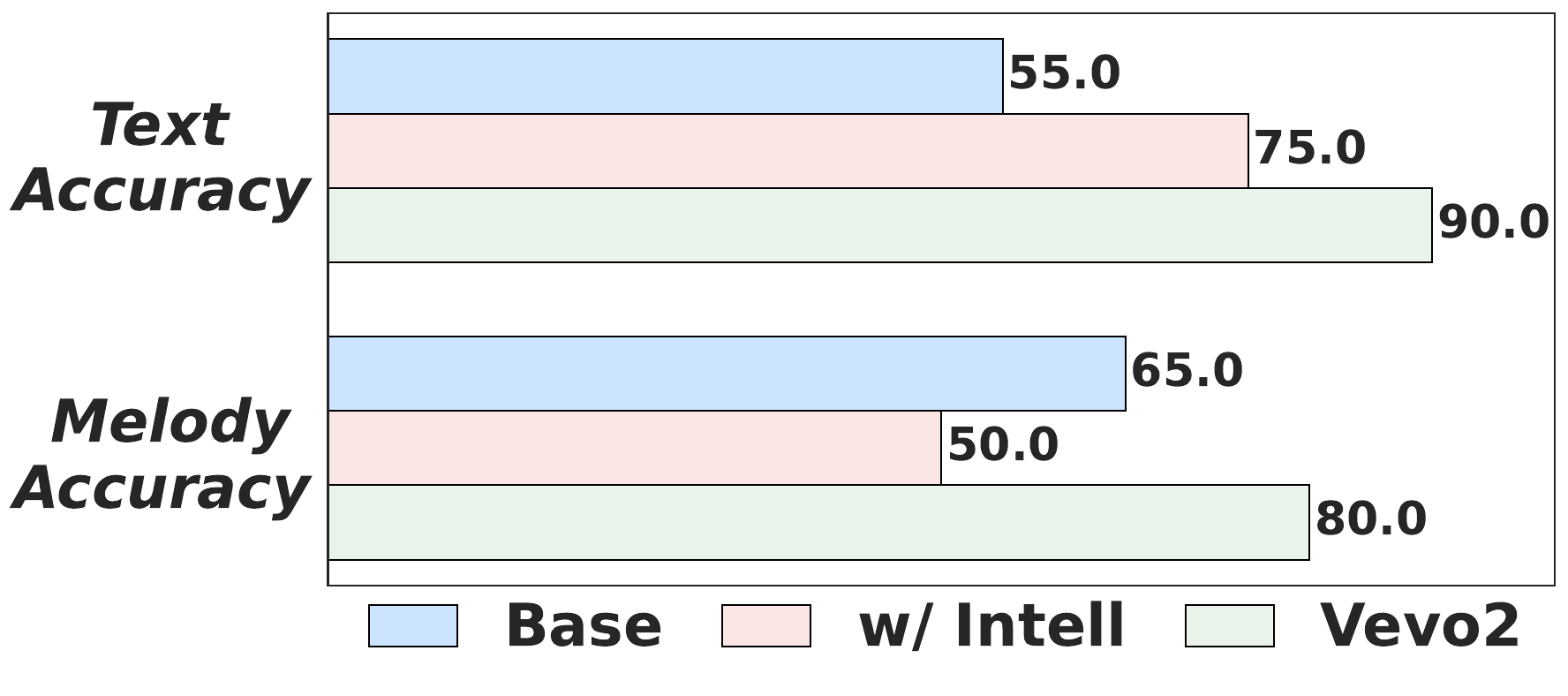}
\end{minipage}
\caption{Effect of intelligibility reward (\textit{w/ Intell}) and prosody similarity reward (\textit{w/ Prosody}) for post-training. The right figure presents subjective evaluation results on the instrument-to-singing task.}
\vspace{-2mm}
\label{fig:grpo-ablation-study}
\end{figure*}

\subsection{Controllability over Pitch Region}\label{sec:expt-controllability-pitch}
The pitch region control is particularly effective for voice conversion (VC) and singing voice conversion (SVC) tasks~\cite{amphion-svc,svcc-vits-ziqian,svcc-2023}. For \vevo{}, to achieve the pitch region control:
\begin{itemize}[itemsep=0ex,leftmargin=3ex]
    \item In the style-preserved VC or SVC tasks, where only the FM model (\vevo{}-FM) is utilized for conversion as shown in Figure~\ref{fig:vevo2-pipeline}: [\colortwo{$\pmb{\bm{cs}}$}, \colortwo{$\pmb{\bm{mel}}$}, \colorone{$\pmb{\bm{cs}}$}] $\rightarrow$ [\colorpredict{$\pmb{\bm{\hat{mel}}}$}], we can shift the F0 of \colorone{\textbf{the source waveform}} to match the pitch region of \colortwo{\textbf{the timbre reference}} before extracting chromagram features. This process yields pitch-region-controlled \colorone{$\pmb{\bm{cs}}$} (while the whisper features for extraction can remain from the original source waveform). 
    \item In the style-converted SVC task, as shown in Figure~\ref{fig:vevo2-pipeline}: [\colortwo{$\pmb{\bm{t}}$}, \colorone{$\pmb{\bm{t}}$}, \colortwo{$\pmb{\bm{p}}$}, \colorone{$\pmb{\bm{p}}$}, \colortwo{$\pmb{\bm{cs}}$}] $\rightarrow$ [\colorpredict{$\pmb{\bm{\hat{cs}}}$}], we similarly shift the F0 of \colorone{\textbf{the source waveform}} to align with \colortwo{\textbf{the timbre reference}}'s pitch region before extracting \colorone{$\pmb{\bm{p}}$}. 
\end{itemize}


\begin{table}[t]
\caption{Effect of pitch region control of \vevo{}.}
\vspace{-1mm}
\label{tab:expt-pitch-control}
\begin{center}
\resizebox{0.5\textwidth}{!}{
\begin{threeparttable}
    \begin{tabular}{l|ccc|ccc}
        \toprule
        \multicolumn{1}{c|}{\multirow{2}{*}{\textbf{Model}}}
        & \multicolumn{3}{c|}{\textbf{English}} 
        & \multicolumn{3}{c}{\textbf{Chinese}} \\
        \cmidrule(lr){2-4} \cmidrule(lr){5-7}
        & \textbf{WER} & \textbf{SIM} & \textbf{FPC}
        & \textbf{WER} & \textbf{SIM} & \textbf{FPC} \\
        \midrule
        \midrule
        \multicolumn{7}{c}{\textbf{\textit{Voice Conversion}}} \\
        \midrule \midrule
        \vevo{}-FM
            & 9.35 & 0.645 & 0.635
            & 6.88 & 0.725 & 0.746 \\
        \quad\textit{w/o pitch shift}
            & 9.07 & 0.607 & 0.628
            & 6.60 & 0.693 & 0.734 \\
        \midrule \midrule
        \multicolumn{7}{c}{\textbf{\textit{Singing Voice Conversion}}} \\
        \midrule \midrule
        \vevo{}-FM
            & 29.82 & 0.587 & 0.826
            & 22.54 & 0.611 & 0.878 \\
        \quad\textit{w/o pitch shift}
            & 27.79 & 0.538 & 0.802
            & 20.58 & 0.545 & 0.876 \\
        \midrule
        \vevo{}
            & 11.64 & 0.601 & 0.782
            & 14.53 & 0.623 & 0.846 \\
        \quad\textit{w/o pitch shift}
            & 11.67 & 0.556 & 0.750
            & 13.15 & 0.563 & 0.817 \\
        \bottomrule
    \end{tabular}
    \vspace{-4mm}
\end{threeparttable}
}
\end{center}
\end{table}

The effectiveness of \vevo{}'s pitch shift control in VC and SVC tasks is demonstrated in Table~\ref{tab:expt-pitch-control}. The results reveal two key findings: First, using pitch shift significantly enhances speaker similarity (SIM) for both \vevo{}-FM (using only the FM model) and the complete \vevo{} system (utilizing both AR and FM models). Second, we observe a slight increase in WER when applying pitch shift. We hypothesize that this minor degradation occurs because the content-style tokens and prosody tokens extracted from inference-time pitch-shifted audio represent patterns less frequently encountered in the training data. In future work, we plan to incorporate such pitch shift perturbations during training to mitigate these small discrepancies arising from the train-inference mismatch.

\subsection{Effectiveness of Multi-Objective Post-Training}\label{sec:ablation-grpo}

To explore the individual effects of intelligibility alignment and prosody similarity alignment, we conduct an ablation study on \vevo{}'s post-training process. Specifically, based on the pre-trained model (i.e., \vevo{}-base), we introduce a post-trained model that exclusively uses intelligibility reward to optimize, denoted as \textit{Base w/ Intell}. We use two challenging tasks to evaluate: \textbf{humming-to-singing} and \textbf{instrument-to-singing}, which also demonstrate \vevo{}'s unique capabilities in prosody control. For the instrument-to-sing task, we design two subjective metrics (see the supplementary materials for details): given a target text, a target melody (i.e., the instrument sound), and a generated voice, subjects perform binary classification to determine whether the voice accurately sings the text (\textit{text accuracy}) and the melody (\textit{melody accuracy}). 

The evaluation sets for humming-to-singing and instrument-to-singing tasks are constructed through the following procedure. For melody sources, we selected humming samples~\cite{humtrans} and instrumental sounds, including flute~\cite{flute-dataset} and violin~\cite{violin-dataset} recordings. For each melody sample, we leveraged its MIDI annotation's note count to generate appropriate lyrics using DeepSeek-v3. The following prompt was used for DeepSeek-v3: \textit{"I will provide the number of notes in a melody. Please write suitable lyrics for it. Note: (1) The number of words in the lyrics should roughly match the number of notes; (2) Please only return the result, without any further explanation."} After obtaining both melody and lyrics, we randomly assigned a target singer from SingStyle111~\cite{singstyle111} for each sample, thereby creating evaluation triplets in the form of (melody, lyric, singer). The evaluation set comprises 1,000 samples in total, evenly distributed between English (500) and Chinese (500) lyrics.

The results shown in Figure~\ref{fig:grpo-ablation-study} reveal several key findings. When using only the intelligibility reward, the model demonstrates significant improvement in intelligibility (WER and Text Accuracy). However, this single-objective optimization potentially compromises other objectives, as evidenced by \textit{Base w/ Intell}'s Melody Accuracy dropping even below that of the pre-trained model (from 65.0\% to 50.0\%). Interestingly, when employing both rewards jointly, we observe both the enhanced melody-following and text-following ability (e.g., text accuracy increases from 75.0\% to 90.0\%). We hypothesize that this improvement stems from strengthened prosody modeling, which in turn benefits pronunciation and consequently enhances intelligibility. This further verifies the advantages of our proposed multi-objective post-training.

\section{Conclusion}

This paper introduces \vevo{}, a framework that bridges controllable speech and singing voice generation through unified prosody learning. Our approach leverages two proposed tokenizers—a notation-free prosody tokenizer that captures prosody and melody from diverse audio sources, and a low-frame-rate and timbre-disentangled content-style tokenizer—along with unified pre-training strategies and a multi-objective post-training phase. \vevo{} demonstrates effectiveness and versatile controllability, achieving strong generalization across a wide array of synthesis, conversion, and editing tasks for both speech and singing, and uniquely enabling novel applications like humming-to-singing and instrument-to-singing. These findings verify the mutual benefits of unified modeling for both speech and singing voice, highlighting promising directions for future research in further unifying diverse audio domains and advancing multi-faceted controllability.

\section*{Acknowledgment}
This work is partially supported by {the National Natural Science Foundation of China under Grant 62376237, 2023 Shenzhen Stability Science Program, Internal Project Fund from Shenzhen Research Institute of Big Data (Grant No. T00120230002), and the Program for Guangdong Introducing Innovative and Enterpreneurial Teams (Grant No. 2023ZT10X044)}. We thank Yicheng Gu, Yushun Zhang and the anonymous reviewers for their insightful comments and suggestions. We appreciate the efforts of all the subjects during the subjective evaluation.

\bibliographystyle{IEEEtran}
\bibliography{ref}

\clearpage
\appendices



\section{Details of Baselines}

\paragraph{Zero-Shot Text to Speech}

\begin{itemize}[itemsep=1ex,leftmargin=3ex]
    \item \textbf{F5-TTS}~\cite{f5tts}: It adopts a single-stage flow-matching transformer to predict Mel spectrograms from text. It is pre-trained on Emilia~\cite{emilia} (101K hours of speech data). We use the officially released checkpoint\footnote{\url{https://huggingface.co/SWivid/F5-TTS}} to generate samples.
    \item \textbf{MaskGCT}~\cite{maskgct}: It adopts a two-stage pipeline: text to content-style tokens (50 Hz), and content-style tokens to acoustic tokens. It is pre-trained on Emilia~\cite{emilia} (101K hours of speech data). We use the officially released checkpoint\footnote{\url{https://huggingface.co/amphion/MaskGCT}} to generate samples.
    \item \textbf{CosyVoice 2}~\cite{cosyvoice}: It contains an AR transformer to predict the content-style tokens (25 Hz) from text, and a flow-matching transformer to predict Mel spectrograms. We use the officially released checkpoint\footnote{\url{https://huggingface.co/FunAudioLLM/CosyVoice2-0.5B}} to generate samples, which is pre-trained on 167K hours of private data.
\end{itemize}

\paragraph{Zero-Shot Singing Voice Synthesis}

\begin{itemize}[itemsep=1ex,leftmargin=3ex]
    \item \textbf{StyleSinger}~\cite{stylesinger}: It is a zero-shot SVS model for style transfer that is pre-trained on the Chinese GTSinger data~\cite{gtsinger}. We use the officially released checkpoint\footnote{\url{https://github.com/AaronZ345/StyleSinger}} to generate samples.
    \item \textbf{TCSinger}~\cite{tcsinger}: It is a zero-shot SVS model for style transfer across cross-lingual speech and singing styles, along with multi-level style control. It is pre-trained on the GTSinger data~\cite{gtsinger}. We use the officially released checkpoint\footnote{\url{https://github.com/AaronZ345/TCSinger}} to generate samples.
\end{itemize}

\paragraph{Zero-Shot Editing}

\begin{itemize}[itemsep=1ex,leftmargin=3ex]
    \item \textbf{SSR-Speech}~\cite{ssrspeech}: It is an AR-based speech editing and synthesis model. We use the officially released Chinese checkpoint\footnote{\url{https://huggingface.co/westbrook/SSR-Speech-Mandarin}} and English checkpoint\footnote{\url{https://huggingface.co/westbrook/SSR-Speech-English}} to generate samples.
    \item \textbf{F5-TTS}~\cite{f5tts}: We use the same pre-trained model as that in the zero-shot TTS task. We use the officially released speech editing script\footnote{\url{https://github.com/SWivid/F5-TTS/blob/main/src/f5_tts/infer/speech_edit.py}} to generate samples.
\end{itemize}

\begin{table*}[t]
\caption{Results of zero-shot TTS task on regular speech domain. The results on expressive speech and singing voice domains are shown in Table~\ref{tab:expt-tts}. (\textbf{PT?}: Whether post-training is applied.)}
\label{tab:expt-tts-regular}
\begin{center}
\resizebox{0.9\textwidth}{!}{
\begin{threeparttable}
    \begin{tabular}{l|c|c|c|rcc|rcc}
        \toprule
        \multicolumn{1}{c|}{\multirow{2}{*}{\makecell[c]{ \textbf{Model}} }} 
        & \multicolumn{1}{c|}{\multirow{2}{*}{\makecell[c]{ \textbf{\#Hz}}}}
        & \multicolumn{1}{c|}{\multirow{2}{*}{\makecell[c]{ \textbf{Data}\\ \textit{(\#hours)}}}}
        & \multicolumn{1}{c|}{\multirow{2}{*}{\makecell[c]{ \textbf{PT?} }}} 
        & \multicolumn{3}{c|}{\textbf{SeedTTS \textit{test-en}}} & \multicolumn{3}{c}{\textbf{SeedTTS \textit{test-zh}}} \\
        \cmidrule(lr){5-7} \cmidrule(lr){8-10}
        & & & & \textbf{WER} & \textbf{SIM} & \textbf{UTMOS} & \textbf{WER} & \textbf{SIM} & \textbf{UTMOS} \\
        \midrule
        \text{Ground Truth}
            & - & - & -
            & 2.143 & 0.730 & 3.523 & 1.254 & 0.750 & 2.793 \\
        \midrule
        \text{F5-TTS}~\cite{f5tts}
            & - & \colorone{\text{101K}} & \ding{55}
            & 3.015 & 0.629 & 3.508 & 3.868 & 0.710 & 2.671 \\
        \text{MaskGCT}~\cite{maskgct}
            & \text{50} & \colorone{\text{101K}} & \ding{55}
            & \textbf{2.403} & \textbf{0.711} & 3.590 & \underline{2.284} & \textbf{0.765} & 2.682 \\
        \text{CosyVoice2}~\cite{cosyvoice}
            & \text{25} & \colorone{\text{167K}} & \ding{51}
            & \underline{2.890} & 0.660 & \textbf{4.100} & \textbf{1.292} & \underline{0.757} & \textbf{3.270} \\
        \midrule
        \multirow{3}{*}{\text{\vevo{}-base}}
            & \multirow{3}{*}{\makecell[c]{\text{12.5}}} & \colorone{\text{101K}} & \ding{55}
            & 5.846 & 0.687 & 3.749 & 7.112 & 0.752 & 3.007 \\
        
            & & \colortwo{\text{7K}} & \ding{55}
            & 42.150 & 0.581 & 3.260 & 16.994 & 0.726 & 2.902 \\
            & & \multirow{1}{*}{\text{\colorone{101K}, \colortwo{7K}}} & \ding{55}
            & 5.770 & 0.686 & 3.763 & 4.632 & 0.752 & \underline{3.009} \\
            \midrule
           \vevo{} & 12.5 & \colorone{101K}, \colortwo{7K} & \ding{51}
            & 3.639 & \underline{0.693} & \underline{3.814} & 2.944 & 0.754 & 2.997 \\
        \bottomrule
    \end{tabular}
    \begin{tablenotes}
        \footnotesize{
            \item[*] \textbf{\#Hz}: Frame rate of the content-style tokenizer. \textbf{Data}: Pre-training data, where \colorone{\textbf{blue}} and \colortwo{\textbf{red}} denote \colorone{\textbf{speech}} and \colortwo{\textbf{singing}} data. The best and the second best result (excluding ground truth) is shown in \textbf{bold} and by \underline{underlined}.
        } 
    \end{tablenotes}
    \vspace{-4mm}
\end{threeparttable}
}
\end{center}
\end{table*}

\paragraph{Zero-Shot Voice Conversion and Singing Voice Conversion}

\begin{itemize}[itemsep=1ex,leftmargin=3ex]
    \item \textbf{FACodec}~\cite{naturalspeech3}: It adopts an auto-encoder and residual vector quantization based architecture. It decouples the raw waveform into factorized attributes through ASR, F0 prediction, and speaker classification tasks, trained on the Libri-light dataset~\cite{libri-light}. We use the released checkpoint in Amphion\footnote{\url{https://huggingface.co/amphion/naturalspeech3\_facodec}}~\cite{amphion,amphion_v0.2} (which is implemented by the authors) to generate samples.
    \item \textbf{Vevo-FM}~\cite{vevo}: It is the flow-matching model of Vevo (i.e., Vevo-Timbre in the original paper), which uses a 50 Hz VQ-VAE content-style tokenizer based on HuBERT~\cite{hubert} representations. It is pre-trained on Emilia (101K hours of speech data). We use the officially released checkpoint\footnote{\url{https://huggingface.co/amphion/Vevo}} to generate samples.
    \item \textbf{CosyVoice2-FM}~\cite{cosyvoice2}: It is the flow-matching model of CosyVoice 2 that uses 25 Hz content-style tokenizer, consistent with the configuration described in the zero-shot TTS task above.
    \item \textbf{NeuCoSVC 2}~\cite{neucosvc}: It is an enhanced version of NeuCoSVC and is pre-trained on an extensive internal dataset comprising approximately 500 hours of singing voice data and various open-source speech datasets~\cite{neucosvc}. We use the officially released checkpoint\footnote{\url{https://github.com/thuhcsi/NeuCoSVC/tree/NeuCoSVC2}} to generate samples.
    \item \textbf{SeedVC}~\cite{seedvc}: It is a diffusion transformer that uses an external timbre shifter during training to perturb the source speech timbre. We use the officially released pre-trained VC and SVC models\footnote{\url{https://github.com/Plachtaa/seed-vc}}, which are denoted as SeedVC (VC) and SeedVC (SVC), to conduct the inference.
\end{itemize}

\paragraph{Zero-Shot Emotion Conversion and Accent Conversion}

\begin{itemize}[itemsep=1ex,leftmargin=3ex]
    \item \textbf{Vevo}~\cite{vevo}: It is named as Vevo-Style in the original paper, which contains an AR transformer to predict content-style tokens (50 Hz) from content tokens, and a flow-matching transformer to predict Mel spectrograms. We use the officially released checkpoint\footnote{\url{https://huggingface.co/amphion/Vevo}} to generate samples, which is pre-trained on 101K hours of Emilia.
\end{itemize}

\section{Additional Experimental Results}\label{sec:appendix-expt}

\begin{revision}

\subsection{Visualization of Different Representations for Modeling Timbre and Melody}
\label{sec:appendix-visualization}

In Section~\ref{sec:tokenizer}, we discretize chromagram features into prosody tokens via a VQ-VAE tokenizer.
This design is motivated by two considerations.
First, discrete tokens naturally fit the sequence modeling paradigm of auto-regressive (AR) transformers, enabling unified conditioning with text and other discrete representations.
Second, we hypothesize that vector quantization can serve as an \emph{information bottleneck}~\cite{autovc,vevo,anyaccomp}: it encourages the model to compress salient melodic/prosodic cues into a compact discrete space, while suppressing residual acoustic characteristics (e.g., timbre or instrument-dependent patterns) that may still be present in continuous chromagram features.

To examine the above hypothesis, we conduct a controlled visualization experiment based on M4Singer~\cite{m4singer}.
We randomly select a set of MIDI annotations (i.e., melodies) and render each MIDI into multiple instrumental audio tracks using pretty\_midi\footnote{\url{https://github.com/craffel/pretty-midi}} with different General MIDI instrument programs.
We also include the original vocal utterance when available.
This construction yields a controlled collection where the \emph{melody is fixed} while the \emph{timbre varies} across instruments.

For each audio sample in this collection, we extract and compare multiple representations: (1) the Mel spectrogram, (2) the chromagram features, and (3) our VQ-quantized chromagram tokens produced by the proposed prosody tokenizer.
To obtain an utterance-level embedding for visualization, we apply mean pooling over time for Mel spectrogram and chromagram features. For VQ tokens, we first map each token index to its corresponding codebook embedding vector, and then mean-pool the resulting embedding sequence over time. Finally, we apply PCA to project all utterance-level embeddings into 2D for visualization.
The results are shown in Figure~\ref{fig:visualization-timbre} and~\ref{fig:visualization-melody}.

\begin{figure}[t]
    \centering

    \begin{minipage}{\columnwidth}
        \centering
        \includegraphics[width=\textwidth]{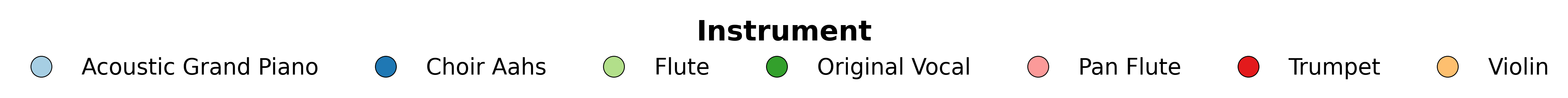}
    \end{minipage}


    \begin{subfigure}{0.32\columnwidth}
        \centering
        \includegraphics[width=\textwidth]{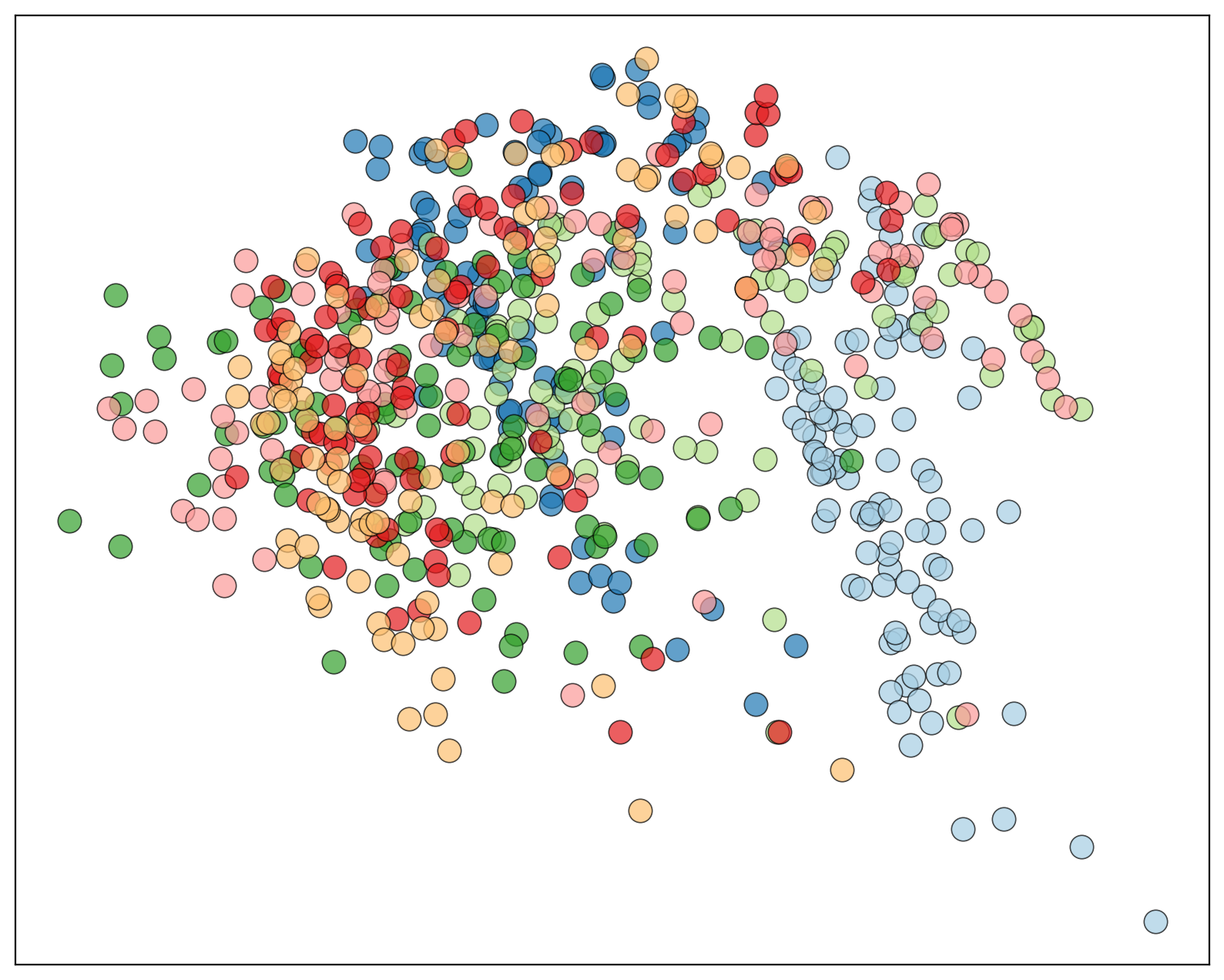}
        \caption{Mel Spectrogram}
        \label{fig:visualization-timbre-mel}
    \end{subfigure}
    \hfill
    \begin{subfigure}{0.32\columnwidth}
        \centering
        \includegraphics[width=\textwidth]{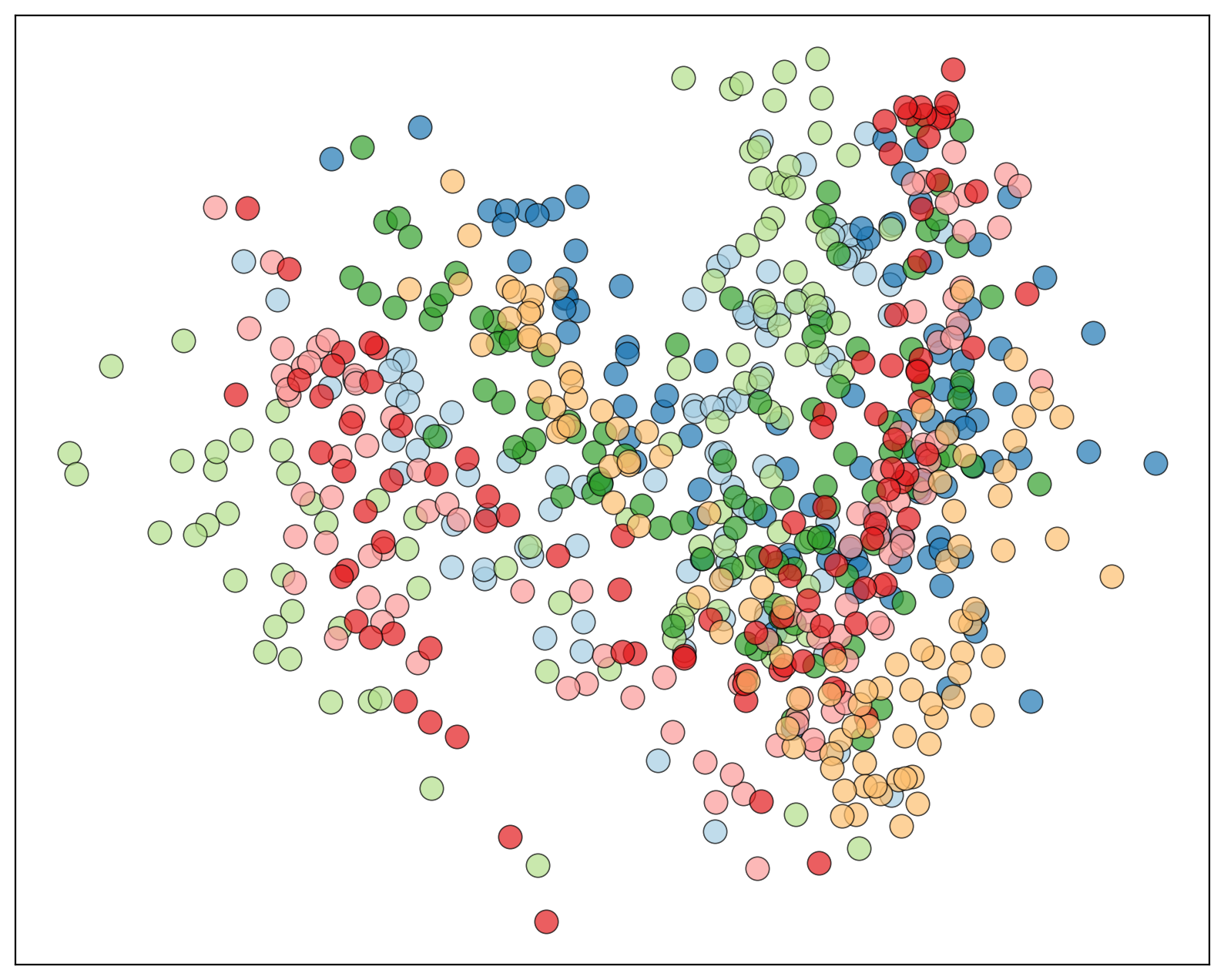}
        \caption{Chromagram}
        \label{fig:visualization-timbre-chroma}
    \end{subfigure}
    \hfill
    \begin{subfigure}{0.32\columnwidth}
        \centering
        \includegraphics[width=\textwidth]{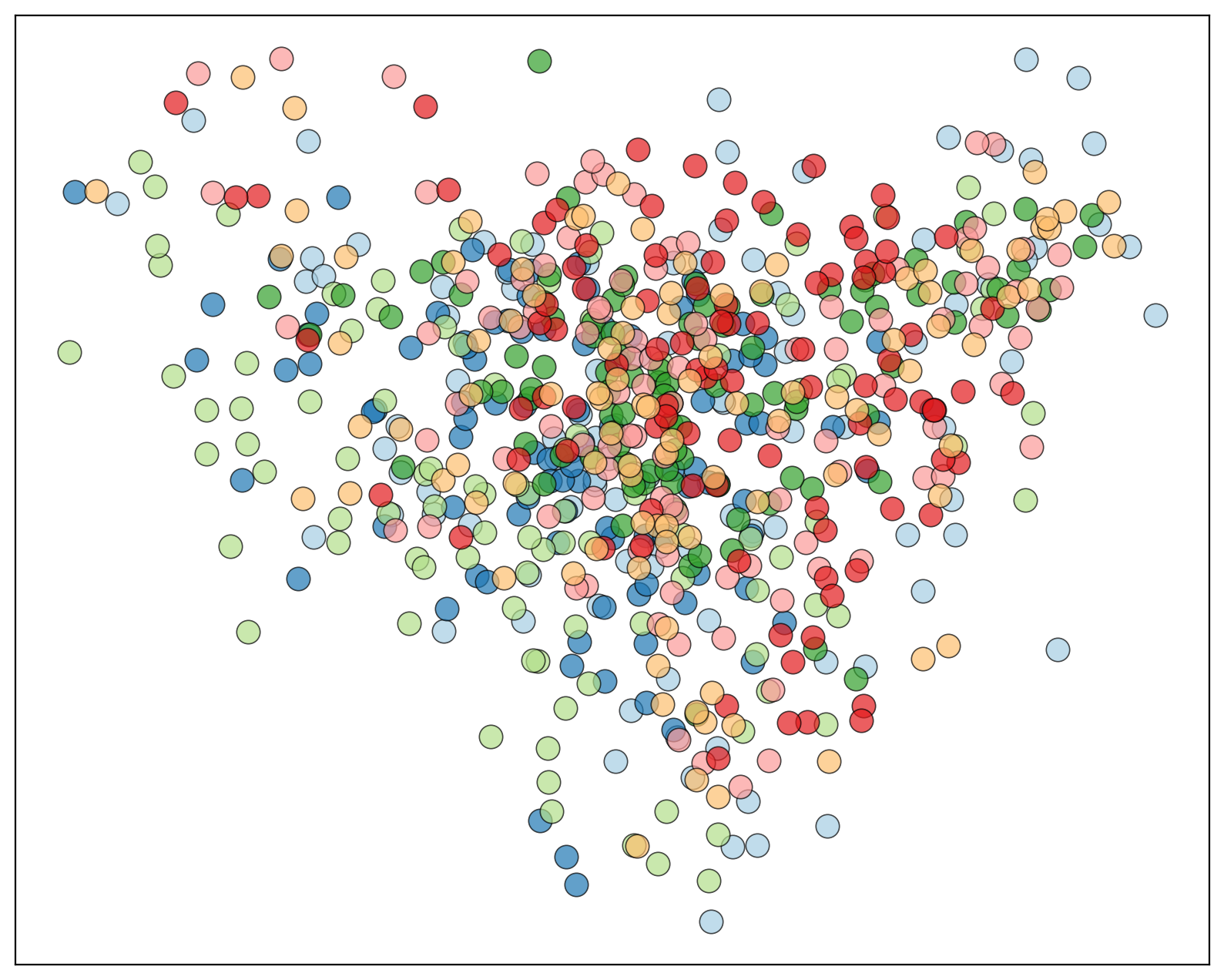}
        \caption{Prosody Token}
        \label{fig:visualization-timbre-chroma-vq}
    \end{subfigure}

    \caption{Timbre invariance of different representations.}
    \vspace{-4mm}
    \label{fig:visualization-timbre}
\end{figure}

\begin{figure}[t]
    \centering
    
    \begin{minipage}{\columnwidth}
        \centering
        \includegraphics[width=\textwidth]{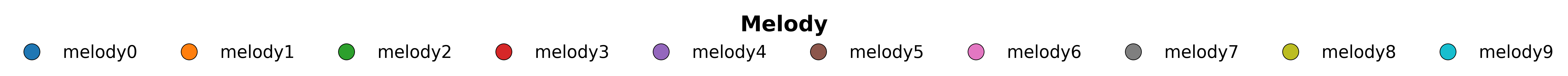}
    \end{minipage}


    \begin{subfigure}{0.32\columnwidth}
        \centering
        \includegraphics[width=\textwidth]{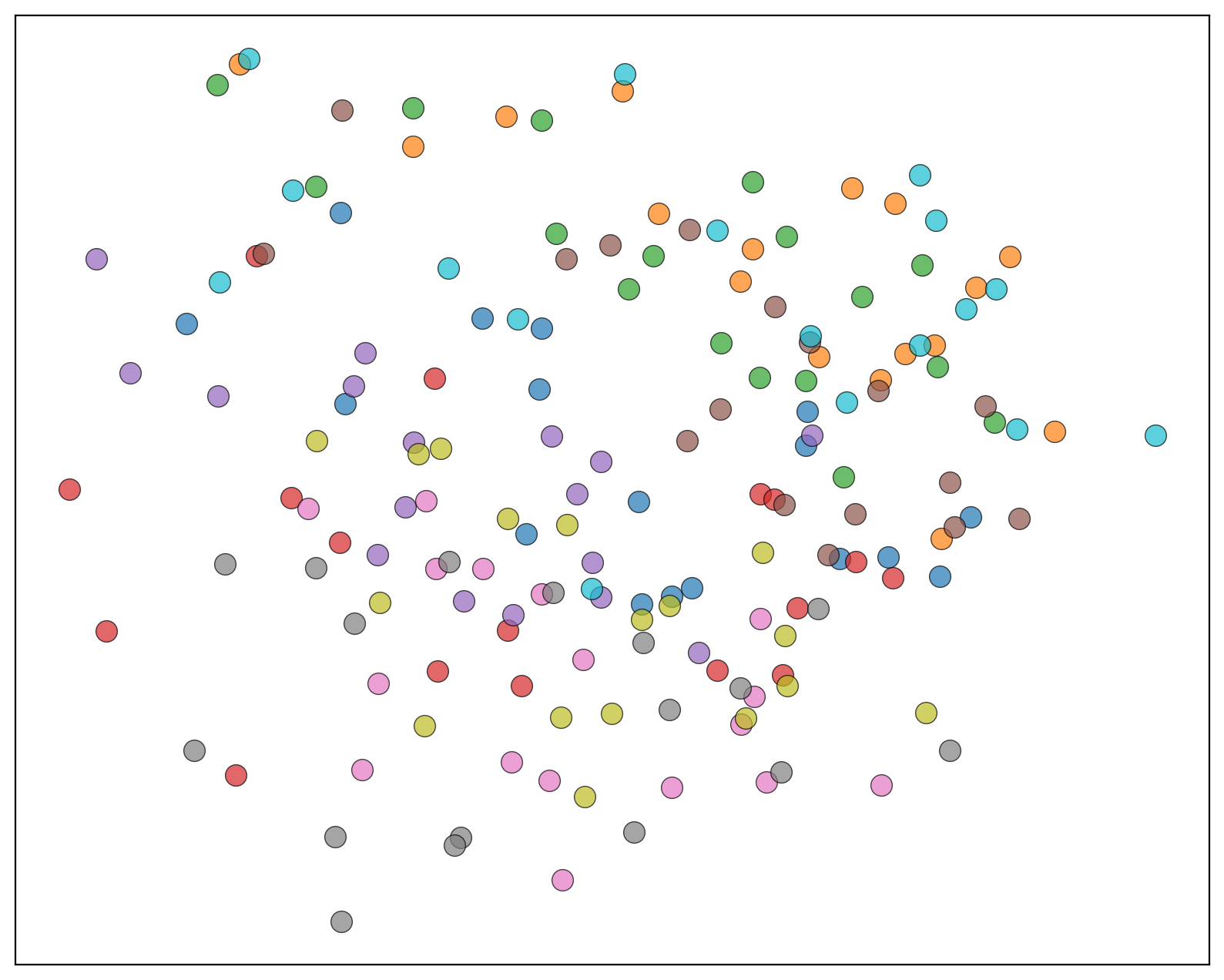}
        \caption{Mel Spectrogram}
        \label{fig:visualization-melody-mel}
    \end{subfigure}
    \hfill
    \begin{subfigure}{0.32\columnwidth}
        \centering
        \includegraphics[width=\textwidth]{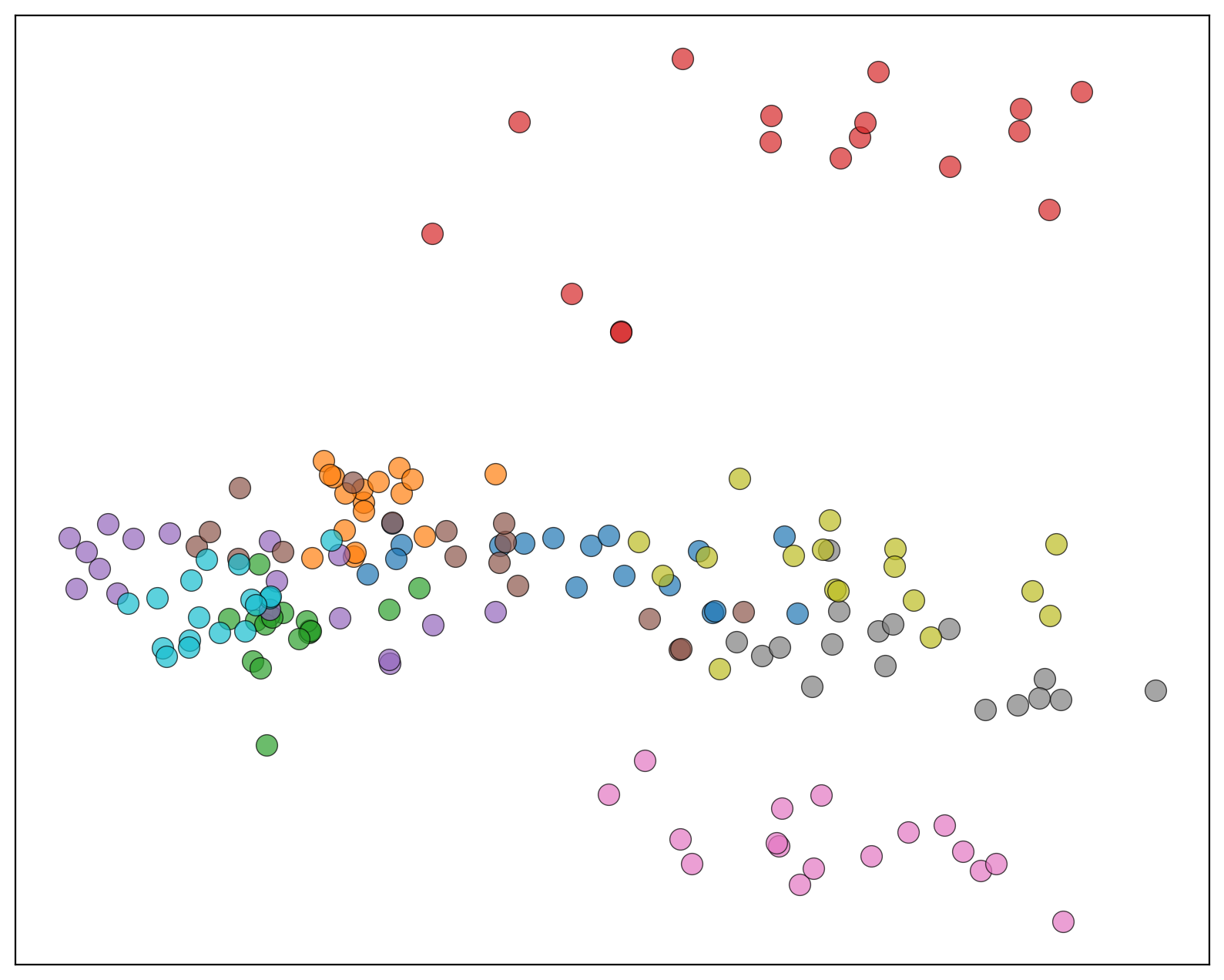}
        \caption{Chromagram}
        \label{fig:visualization-melody-chroma}
    \end{subfigure}
    \hfill
    \begin{subfigure}{0.32\columnwidth}
        \centering
        \includegraphics[width=\textwidth]{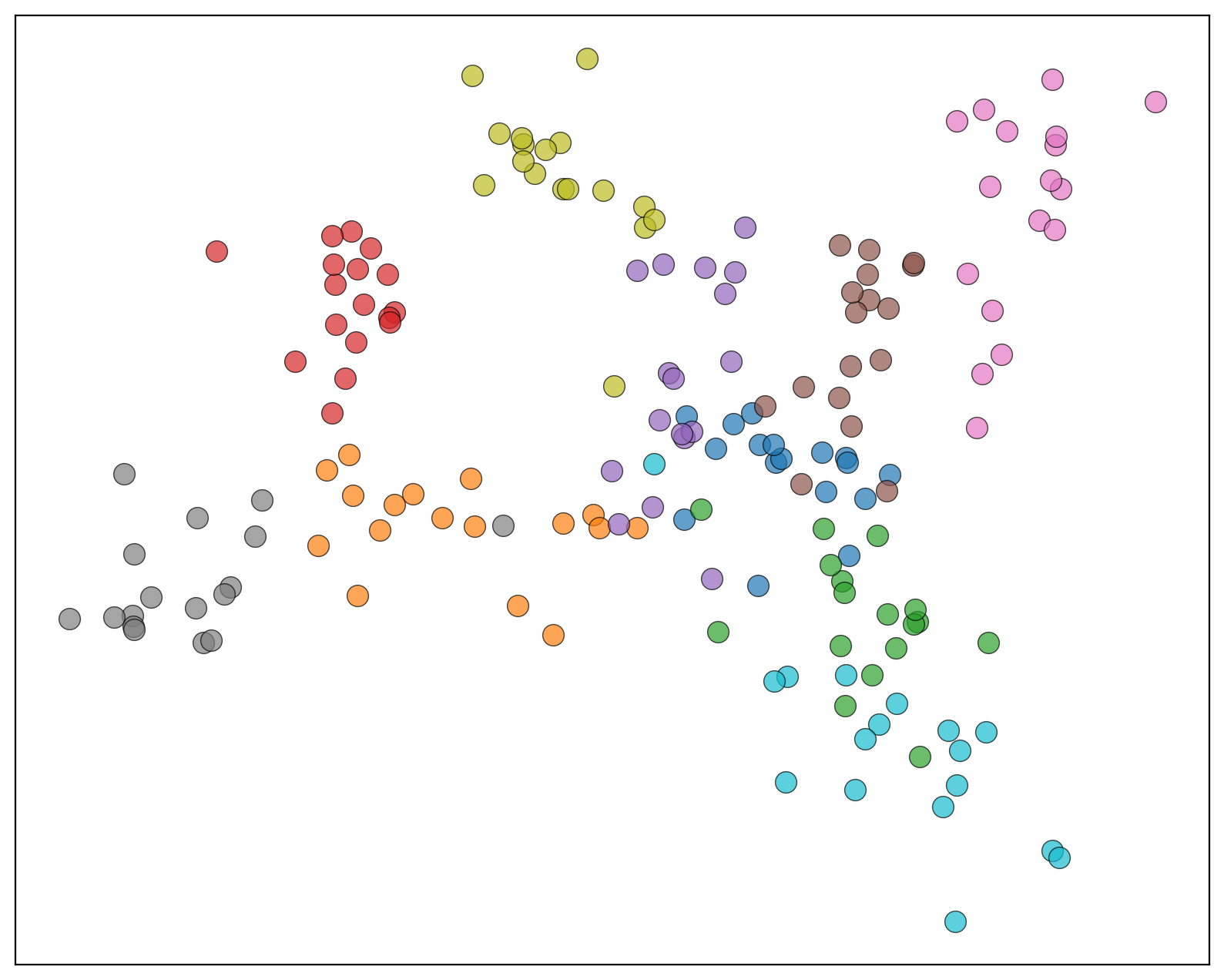}
        \caption{Prosody Token}
        \label{fig:visualization-melody-chroma-vq}
    \end{subfigure}

    \caption{Melodic clusterability of different representations.}
    \vspace{-4mm}
    \label{fig:visualization-melody}
\end{figure}

We highlight two properties that are desirable for a prosody/melody representation in controllable voice generation:

\textbf{(1) Timbre invariance.}\quad
A robust prosody representation should be insensitive to timbre so that it generalizes across speakers and instruments.
In Figure~\ref{fig:visualization-timbre}, the Mel spectrogram (Figure~\ref{fig:visualization-timbre-mel}) exhibits noticeable clustering by instrument type, indicating strong timbre dependence. In contrast, the dense chromagram (Figure~\ref{fig:visualization-timbre-chroma}) already mitigates this bias by substantially mixing samples from different instruments.
Notably, the VQ-quantized chromagram, i.e., our proposed prosody token (Figure~\ref{fig:visualization-timbre-chroma-vq}), further increases the mixing across instruments, suggesting that discretization helps reduce residual timbre cues and promotes a more timbre-invariant representation.

\textbf{(2) Melodic clusterability.}\quad
An effective prosody representation should map identical melodies to compact regions in the embedding space regardless of timbre variations.
Figure~\ref{fig:visualization-melody} shows that Mel spectrogram (Figure~\ref{fig:visualization-melody-mel}) fail to form coherent clusters by melody ID, likely due to their entanglement with diverse acoustic factors.
Dense chromagram features (Figure~\ref{fig:visualization-melody-chroma}) yield clearer melodic grouping, indicating improved melody sensitivity.
Moreover, the VQ-quantized chromagram, i.e., our proposed prosody token (Figure~\ref{fig:visualization-melody-chroma-vq}), produces more compact and better-separated clusters, implying that quantization preserves melodic structure while providing a robust discrete conditioning signal for controllable generation.

\end{revision}

\subsection{Zero-Shot Text to Speech}\label{sec:appendix-expt-tts}

The zero-shot TTS evaluation results for expressive speech and singing voice domains are presented in Table~\ref{tab:expt-tts}. Table~\ref{tab:expt-tts-regular} demonstrates the performance on the regular speech domain using the SeedTTS evaluation sets~\cite{seedtts}, a widely adopted benchmark in the TTS field~\cite{maskgct,cosyvoice2,f5tts}. The results corroborate our findings from Section~\ref{sec:expt-mutual-benefits}. Specifically, we observe that incorporating singing voice data alongside speech data enhances performance across all metrics (WER, SIM, UTMOS) in the regular speech domain. Notably, the post-trained \vevo{} achieves comparable performance to existing state-of-the-art baselines, despite utilizing a content-style tokenizer with a lower frame rate.






\subsection{Zero-Shot Voice Conversion}\label{sec:appendix-zero-shot-vc}

\begin{table*}[t]
\caption{Results on zero-shot VC task.}
\vspace{-1mm}
\label{tab:expt-vc}
\begin{center}
\resizebox{0.65\textwidth}{!}{
\begin{threeparttable}
    \begin{tabular}{l|ccc|ccc}
        \toprule
        \multicolumn{1}{c|}{\multirow{2}{*}{\textbf{Model}}}
        & \multicolumn{3}{c|}{\textbf{English}} 
        & \multicolumn{3}{c}{\textbf{Chinese}} \\
        \cmidrule(lr){2-4} \cmidrule(lr){5-7}
        & \textbf{WER} & \textbf{SIM} & \textbf{UTMOS}
        & \textbf{WER} & \textbf{SIM} & \textbf{UTMOS} \\
        \midrule
        \text{Ground Truth}
            & 2.15 & - & 3.52
            & 1.25 & - & 2.79 \\
        \midrule
        \text{FACodec}~\cite{naturalspeech3}
            & 4.88 & 0.355 & 2.79
            & 8.20 & 0.495 & 2.10 \\
        \text{Vevo-FM}~\cite{vevo}
            & \underline{3.01} & 0.536 & 3.51
            & 4.06 & 0.695 & 2.86 \\
        \text{CosyVoice2-FM}~\cite{cosyvoice2}
            & 4.66 & 0.530 & \textbf{3.86}
            & \underline{2.80} & 0.728 & \textbf{3.08} \\
        \midrule
        \text{NeuCoSVC 2}~\cite{neucosvc}
            & 3.57 & 0.227 & 3.15
            & 2.90 & 0.468 & 2.51 \\
        \text{SeedVC (VC)}~\cite{seedvc}
            & \textbf{2.97} & 0.565 & 3.31
            & \textbf{2.45} & \underline{0.737} & 2.69 \\
        \midrule
        \text{\vevo{}-FM}
            & 9.35 & \underline{0.645} & 3.59
            & 6.88 & 0.725 & 2.83 \\
        \text{\vevo{}}
            & 3.53 & \textbf{0.692} & \underline{3.81}
            & 3.01 & \textbf{0.755} & \underline{3.00} \\
        \bottomrule
    \end{tabular}
    \begin{tablenotes}
        \footnotesize{
            \item[*] \vevo{}-FM denotes the flow-matching model of \vevo{}. The best and the second best results (excluding ground truth) are shown in \textbf{bold} and \underline{underlined}.
        }
    \end{tablenotes}
\end{threeparttable}
}
\end{center}
\end{table*}

Table~\ref{tab:expt-vc} presents the results of our voice conversion evaluation, utilizing the same baselines\footnote{For SeedVC, we employed the official SVC checkpoint for the zero-shot SVC task (Table~\ref{tab:expt-svc}) and the official VC checkpoint for the zero-shot VC task (Table~\ref{tab:expt-vc}).} as the SVC task in Table~\ref{tab:expt-svc}, where we use the SeedTTS VC evaluation set~\cite{seedtts}. The VC task results lead to conclusions similar to those observed in the SVC experiments: First, \vevo{}-FM demonstrates robust timbre disentanglement capabilities, with speaker similarity (SIM) metrics matching or exceeding existing VC models, particularly for English content. Furthermore, the transition from \vevo{}-FM to \vevo{} (i.e., from style-preserved VC to style-converted VC) yields additional improvements in speaker similarity. Moreover, the explicit incorporation of text input significantly enhances intelligibility performance.




\section{Subjective Evaluation}\label{sec:appendix-subeval}

We recruited multiple subjects on a compensated basis to conduct subjective evaluations. All participants possessed extensive musical backgrounds, with a minimum of five years of experience in instrumental performance and music production. Each audio sample in our evaluation received at least three independent ratings. We have developed an automated subjective evaluation interface. For each item to be evaluated, users will see three components: the System Interface (i.e., the audio to be evaluated), the Questionnaire, and the Evaluation Criteria.

\paragraph{Naturalness (N-CMOS)}

\begin{itemize}[itemsep=0ex,leftmargin=2ex]
    \item \textbf{System Interface:} Users listen to two speech samples, A and B, to compare their naturalness.
    \item \textbf{Questionnaire:} Which speech sample sounds more natural and human-like, rather than synthetic or AI-generated?
    \item \textbf{Evaluation Criteria:} Options include A +2 (Sample A is much more natural), A +1 (Sample A is slightly more natural), Tie (Both are equally natural), B +1 (Sample B is slightly more natural), and B +2 (Sample B is much more natural).
\end{itemize}

\paragraph{Speaker Similarity (SS-CMOS)}

\begin{itemize}[itemsep=0ex,leftmargin=2ex]
    \item \textbf{System Interface:} Users listen to two speech samples, A and B, to evaluate their similarity to the speech of the reference speaker.
    \item \textbf{Questionnaire:} Which speech sounds more like the reference speaker's voice?
    \item \textbf{Evaluation Criteria:} Options include A +2 (Sample A is much more similar), A +1 (Sample A is slightly more similar), Tie (Both are equally similar), B +1 (Sample B is slightly more similar), and B +2 (Sample B is much more similar).
\end{itemize}

\paragraph{Prosody Similarity (PS-CMOS)}

\begin{itemize}[itemsep=0ex,leftmargin=2ex]
    \item \textbf{System Interface:} Users listen to two speech samples, A and B, to evaluate their similarity to the speech of the reference prosody.
    \item \textbf{Questionnaire:} Ignore the vocal characteristics (who is speaking), just pay attention to the speaking style (how to speak). Which speech's speaking style (pace, tone, stress, intonation, pitch, emotion) is more consistent and identical with the reference voice?
    \item \textbf{Evaluation Criteria:} Options include A +2 (Sample A is much more similar), A +1 (Sample A is slightly more similar), Tie (Both are equally similar), B +1 (Sample B is slightly more similar), and B +2 (Sample B is much more similar).
\end{itemize}

\paragraph{Melody Similarity (MS-CMOS)}

\begin{itemize}[itemsep=0ex,leftmargin=2ex]
    \item \textbf{System Interface:} Users listen to two singing voice samples, A and B, to evaluate their similarity to the singing voice of the reference melody.
    \item \textbf{Questionnaire:} Which singing voice's melody is more consistent and identical with the reference voice?
    \item \textbf{Evaluation Criteria:} Options include A +2 (Sample A is much more similar), A +1 (Sample A is slightly more similar), Tie (Both are equally similar), B +1 (Sample B is slightly more similar), and B +2 (Sample B is much more similar).
\end{itemize}

\paragraph{Style Similarity (Style-CMOS)}

\begin{itemize}[itemsep=0ex,leftmargin=2ex]
    \item \textbf{System Interface:} Users listen to two singing voice samples, A and B, to evaluate their similarity to the singing voice of the reference style.
    \item \textbf{Questionnaire:} Which singing voice's vocal techniques and style (such as the \textit{vibrato}, \textit{falsetto}, \textit{glissando}, etc.) is more similar to the reference voice?
    \item \textbf{Evaluation Criteria:} Options include A +2 (Sample A is much more similar), A +1 (Sample A is slightly more similar), Tie (Both are equally similar), B +1 (Sample B is slightly more similar), and B +2 (Sample B is much more similar).
\end{itemize}

\paragraph{Melody-MOS}

\begin{itemize}[itemsep=0ex,leftmargin=2ex]
    \item \textbf{System Interface:} One reference audio, One audio to be evaluated.
    \item \textbf{Questionnaire:} Does the singing voice's melody follow the reference audio?
    \item \textbf{Evaluation Criteria:} Options include 1 (Unable to follow), 2 (Roughly following the melody contour), and 3 (Completely following all melody details).
\end{itemize}

\paragraph{Text Accuracy}
    
\begin{itemize}[itemsep=0ex,leftmargin=2ex]
    \item \textbf{System Interface:} Users listen to the singing voice and compare it to the provided target text to assess whether the voice matches the text.
    \item \textbf{Questionnaire:} Does the voice contain any pronunciation errors in the lyrics (such as insertion, omission, or mispronunciation)?
     \item \textbf{Evaluation Criteria:} The evaluation is binary: ``No Error'' (the voice matches the text) or ``Has Error'' (the voice does not match the text).
\end{itemize}

\paragraph{Melody Accuracy}
    
\begin{itemize}[itemsep=0ex,leftmargin=2ex]
    \item \textbf{System Interface:} Users listen to the singing voice and compare it to the provided target humming or instrumental melody to assess whether the voice matches the melody.
    \item \textbf{Questionnaire:} Does the voice follow the target melody without any out-of-tone errors?
     \item \textbf{Evaluation Criteria:} The evaluation is binary: ``No Error'' (the voice matches the melody) or ``Has Error'' (the voice does not match the melody).
\end{itemize}

\end{document}